\documentclass{article}

\usepackage{PRIMEarxiv}

\usepackage{amsmath}
\usepackage{subcaption}
\usepackage{algorithm,algorithmic}
\usepackage[utf8]{inputenc} 
\usepackage[T1]{fontenc}    
\usepackage{hyperref}       
\usepackage{url}            
\usepackage{booktabs}       
\usepackage{amsfonts}       
\usepackage{nicefrac}       
\usepackage{microtype}      
\usepackage{lipsum}
\usepackage{fancyhdr}       
\usepackage{graphicx}       

\usepackage[letterpaper]{geometry}
\usepackage{booktabs}
\usepackage{placeins}
\usepackage{newfloat}
\usepackage{listings}
\usepackage{bbm}
\usepackage{bm}
\usepackage{appendix}

\graphicspath{{media/}}     

\title{Hierarchical-Graph-Structured Edge Partition Models for Learning Evolving Community Structure
}

\author{
  Xincan Yu \\
  Great Bay Institute for Advanced Study \\
  Great Bay University \\
  Guangdong, China\\
  \texttt{xincanyu385@gmail.com} \\ 
   \And
  Sikun Yang \thanks{Corresponding author} \\
  School of Computing and Information Technology \\
  Great Bay University \\
  Guangdong, China\\
  \texttt{sikunyang@gbu.edu.cn} \\
}

\begin{document}
\maketitle

\begin{abstract}
We propose a novel dynamic network model to capture evolving latent communities within temporal networks. To achieve this, we decompose each observed dynamic edge between vertices using a Poisson-gamma edge partition model, assigning each vertex to one or more latent communities through \emph{nonnegative} vertex-community memberships. Specifically, hierarchical transition kernels are employed to model the interactions between these latent communities in the observed temporal network. A hierarchical graph prior is placed on the transition structure of the latent communities, allowing us to model how they evolve and interact over time. Consequently, our dynamic network enables the inferred community structure to merge, split, and interact with one another, providing a comprehensive understanding of complex network dynamics. Experiments on various real-world network datasets demonstrate that the proposed model not only effectively uncovers interpretable latent structures but also surpasses other state-of-the art dynamic network models in the tasks of link prediction and community detection. 
\end{abstract}


\section{Introduction}
Modelling dynamic network data has been receiving a growing amount of attention, because modelling dynamic network forms the basis of many downstream studies including recommendation system design ~\cite{walter2008model, he2010social, dong2012link}, information cascade prediction ~\cite{li2014deep, lei2019gcn, choudhury2024community}, influence maximisation ~\cite{xie2015dynadiffuse, kumar2022influence}, network causal inference ~\cite{smith2018influence, bhattacharya2020causal, ogburn2024causal}, and etc. The earliest study of modelling dynamic networks can be traced back to the dynamic stochastic block model ~\cite{holland1983stochastic, nowicki2001estimation, matias2017statistical}. Kemp et al. \cite{kemp2006learning} assume a dynamic network consists of several latent communities, and assigns each vertex to one of the latent communities, using a discrete-distributed vertex-community membership vector. To capture \emph{overlapping} community structure, Airoldi et al. \cite{airoldi2008mixed} further develop a dynamic mixed membership stochastic block model which allows each vertex to associate with multiple latent communities, using multinomial-distributed vertex-community memberships. Following that success, Heaukulani and Ghahramani  
\cite{heaukulani2013dynamic} propose latent feature propagation that captures the affiliation of each vertex to the multiple latent communities, using \emph{binary} memberships. However, this approach does not account for the differences in the degrees of the memberships of each vertex to the latent communities. To further enhance model capability, Yang and Koeppl \cite{yang2018poisson,DBLP:conf/icml/YangK18} propose a dynamic Poisson gamma membership model, which captures network dynamics using gamma-distributed vertex-community memberships, and hence admits greater flexibility in representing complicated community structure, in comparisons to prior probabilistic dynamic network models ~\cite{zhou2015infinite, acharya2015nonparametric}. Moreover, some recent work ~\cite{gao2022graph, li2014deep, liu2024community} study deep neural networks-based methods for capturing temporal networks, and demonstrate excellent accuracy in link prediction and community detection. 

Despite achieving many successes in capturing complicated community structures, the aforementioned methods still fail to model how the latent communities evolve and excite with each other, over time. For instance, two small communities may interact with each other, and gradually merge together into a large one. To capture such evolving behaviours of the underlying communities, this paper aims to study a novel dynamic network model, in which each vertex is affiliated to multiple communities with a gamma-distributed vertex-community membership vector. A transition kernel is introduced to capture how those latent communities evolve over time. In particular, we also study a hierarchical prior to impose graph structure over the transition kernel, to enhance the model explainability in capturing real-world network dynamics. Moreover, the hierarchical prior construction of the proposed dynamic network model, enables us to select an appropriate number of latent communities, which thus avoids the tedious issue of model selection specifically for large temporal networks. 

The main contributions of the work include: 
\begin{itemize}
    \item[$\bullet$] A novel Bayesian model is developed to track how the latent communities evolve and excite with each other, over time. 
    \item[$\bullet$] A hierarchical prior is specifically dedicated to imposing sparse graph structure over the transition kernel, which naturally enables us to capture how the communities interact with each other, and thus enjoys stronger model explainability, in comparisons to previously related probabilistic dynamic network models.
    \item[$\bullet$] The final experiments, conducted on several real-world dynamic network data, show the superior performance of the proposed model in link prediction and community detection, in comparison to the prior state-of-art methodologies. The estimated network parameters demonstrate the great explainability of our novel dynamic network model. 
\end{itemize}

\section{The Proposed Dynamic Network Model}
In this section, we shall first describe the problem setting, and then introduce the details of the novel hierarchical-structured edge partition model for tracking the evolution of underlying communities.

\noindent \textbf{Temporal Network Data.} 
In this context, we only consider temporal networks in which the number of vertices does not change, but the edges between the observed vertices may appear or disappear over time. More specifically, a temporal network sampled over $T$ discrete time steps consists of a sequence of adjacency matrices $\mathbb{B} := \{B^{(t)}\}_{t=1}^{T}$, where $B^{(t)} := [b_{ij}^{(t)}]$ is a binary matrix of size $ \vert \mathbb{V} \vert \times \vert \mathbb{V} \vert$, $\mathbb{V}$ is the collection of the vertices, and $\vert \mathbb{V} \vert$ denotes the number of vertices. More specifically, $b_{ij}^{(t)} = 1$ means that there exists an edge between two vertices $i$ and $j$ at time $t$, and vice versa. The main goal here is, to detect complicated latent structures given the observed network dynamics, and also to infer how those latent communities interact with each other through a transition kernel.

\noindent \textbf{Hierarchical-structured Edge Partition Model.} 
\begin{figure}[hbt!]
    \centering
    \includegraphics[width=1.0\columnwidth]{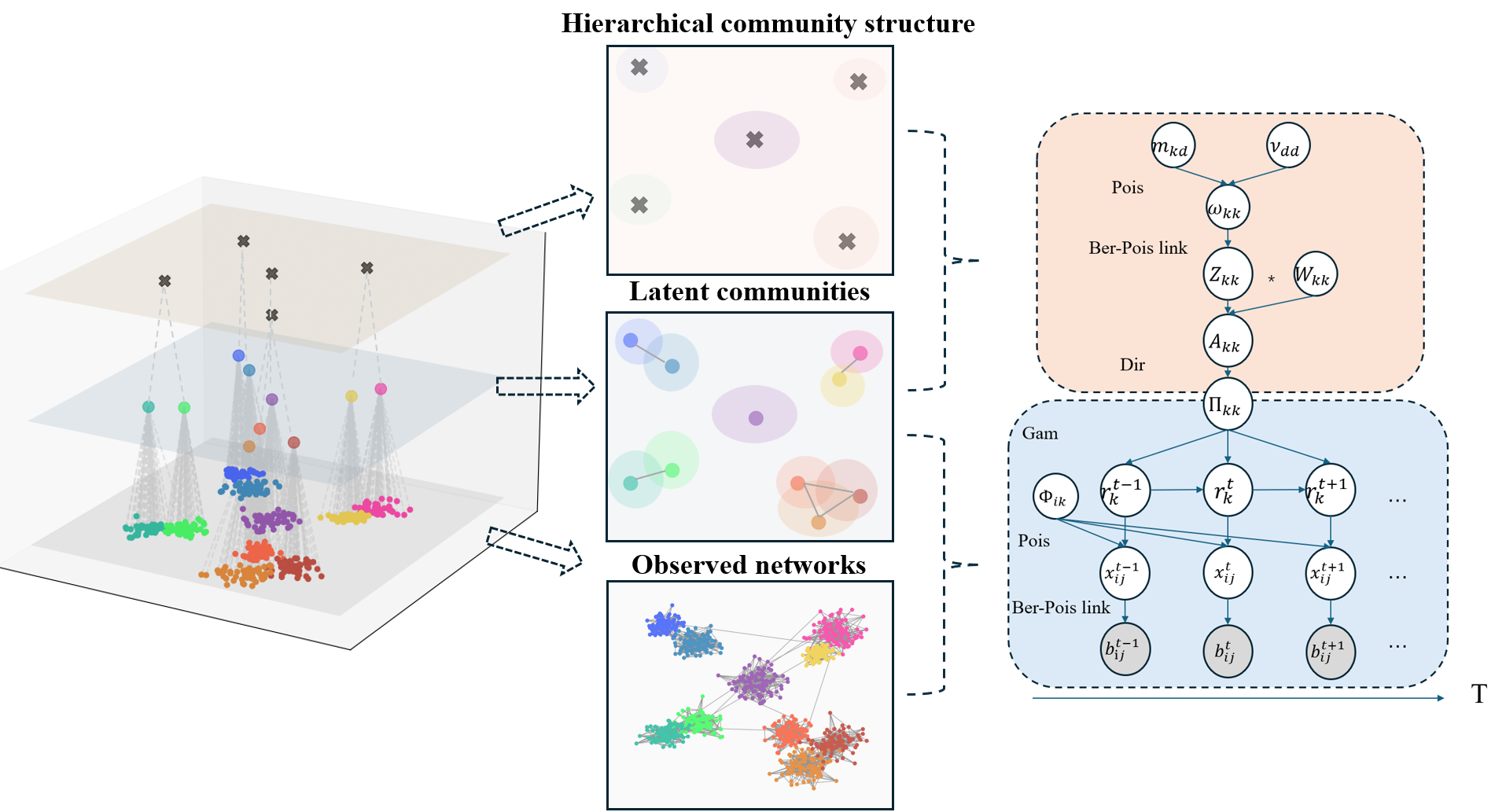}
    \caption{The left plot illustrates the data mapping across the three layers in our model, G-HSEPM. The bottom layer represents the observations, where some vertices are connected by edges. In the middle layer, vertices with strong connections naturally form communities. The top layer further groups these communities into hierarchical clusters based on their interactions. To clarify the relationships within each layer, we decompose the data at each stage in the middle plot. The right plot shows how the model operates across these layers: the blue section captures vertex-level memberships, while the pink section models community-level memberships.}
    \label{Working_Digram}
\end{figure}
We shall first describe how to represent the latent community structure at one time step, and then explain how to capture the evolution behaviours of the latent communities over time. To capture the overlapping community structure, we use a nonnegative vertex-community membership vector, to describe the degrees of each vertex belonging to the multiple latent communities. More specifically, suppose the observed dynamic network consists of $K$ communities, we draw the membership of vertex $i$ belonging to the community $k$ from a gamma distribution as $\phi_{ik} \sim \text{Gam} (a_0, 1/c_i)$, where $a_0$ is a hyperparameter and $c_i$ regularises the degree of vertex $i$. We assume that the latent communities are changing over time, and thus use a time-dependent variable $r_{k}^{(t)}$ to represent the status of community $k$ at time $t$. Intuitively, we draw $r_{k}^{(t)}$ as $r_{k}^{(t)} \sim \text{Gam} (\sum_{k_2=1}^{K} \pi_{kk_2} r_{k_2}^{(t-1)}, 1/\tau) $, for which we can understand that the weight of community $k$ at time $t$, can be affected by the weights of all the communities at last time step $t-1$, and the corresponding influence coefficients are determined by $\Pi_k:=[\pi_{k1},\cdots, \pi_{kk}]^{T}$. In particular, we draw the community weights at an initial time independently from a gamma prior as $r_{k}^{(1)} \sim \text{Gam} (\tau\nu, 1/\tau)$. We impose a Dirichlet prior over the transition kernel as $\pi_k \sim \text{Dir} (\nu_1\nu_k,\cdots,\xi\nu_k,\cdots,\nu_K\nu_k)$, and draw the hyperparameter $\nu_k$ as $\nu_k \sim \text{Gam} (\frac{\gamma_0}{K}, 1/\beta) $, where $\gamma_0$ is the concentration parameter, and $\beta$ is the hyperparameter. Note that as the number of latent communities $K$ goes to infinity, the hierarchical gamma prior will shrink the redundant communities with small weights toward zeros, and thus can effectively choose an appropriate number of latent communities for real-world temporal networks. Given the vertex-community memberships $\bm{\phi}_{i}$, $\bm{\phi}_{j}$, and the community weights  $r^{(t)}_{k}$, we capture the probabilities of the edge between the two vertices $i$ and $j$, using Bernoulli-Poisson link function as 
\begin{equation}
    b_{ij}^{(t)} \sim \text{Bernoulli}(1-\text{exp}(-\sum_{k=1}^{K} r_{k}^{(t)} \phi_{ik} \phi_{jk})).
    \label{ber}
\end{equation}

Interestingly, Eq.(\ref{ber}) can be equivalently represented as 
\begin{align}
    &b_{ij}^{(t)} = \mathbbm{1} (x_{ij}^{(t)} \geq 1), \\
    &x_{ij}^{(t)} \sim \text{Pois}\left(\sum_{k=1}^{K}r_{k}^{(t)} \phi_{ik} \phi_{jk}\right),
\end{align}
which enables us to infer the model parameters in a tractable way (details in Section \ref{Inference}). The full generative hierarchical-structured edge partition (HSEPM) model is specified as  
\begin{alignat}{2}
&{b}_{ij}^{(t)} &&= \mathbbm{1}({x}_{ij}^{(t)}\geq1),  \\
&{x}_{ij}^{(t)} &&\sim \text{Pois}\left(\sum_{k=1}^{K}r_{k}^{(t)} \phi_{ik} \phi_{jk}\right), \\
&\phi_{ik} &&\sim \text{Gam}(a_0, 1/c_{i}), \\
&r_{k}^{(t)} &&\sim \text{Gam}\left(\sum_{k_2=1}^{K} \pi_{kk_2} r_{k_2}^{(t-1)}, 1/\tau\right),\text{for t = 2,} \cdots \text{, T}, \\
&r_{k}^{(1)} &&\sim \text{Gam}\left(\tau\nu, 1/\tau\right), \\
&\pi_k &&\sim \text{Dir}\left(\nu_1\nu_k,\cdots,\xi\nu_k, \cdots, \nu_K\nu_k \right), \\
&\nu_k &&\sim \text{Gam}\left(\frac{\gamma_0}{K}, 1/\beta\right). \\
\end{alignat}
Its graphical representation is presented in Appendix \ref{Gibbs_HSEPM}.

\noindent \textbf{Graph-hierarchical-structured Edge Partition Model.}
To further capture the interaction structure between latent communities, we impose a hierarchical prior to inducing a sparse-graph-structured-transition kernel as 
\begin{align}
    &\pi_k \sim \text{Dir} (\alpha_{1k},\cdots,\alpha_{Kk} ), \\
    &\alpha_{k_1k_2} =  z_{k_1k_2} \cdot w_{k_1k_2},
\end{align}
where $\bm{\alpha}_k = [\alpha_{1k},\cdots, \alpha_{kk}]^{\mathrm T}$ is the hyperparameter of the Dirichlet prior. We construct this hierarchical prior by first drawing the weight parameter $w_{k_1k_2}$ from a gamma distribution as $w_{k_1k_2} \sim \text{Gam} (e_0, e_0)$, and then multiplying it by a binary variable $z_{k_1k_2}$ sampled from a hierarchical graph process as 
\begin{equation}
    z_{k_1k_2} \sim \text{Bernoulli} (1-\text{exp}(-\sum_{d_1=1}^{D} \sum_{d_2=1}^{D} m_{k_1d_1} v_{d_1d_2} m_{k_2d_2})),
\end{equation} 
where we further factorise the $K$ latent communities into $D$ hierarchical communities. This formulation, in other words, captures the latent structure underlying the transition dynamics between communities. Specifically, $m_{kd}$ represents the degree of association between the $k$-th latent community and the $d$-th hierarchical community, while $v_{d_1d_2}$ denotes the compatible weight between the two hierarchical communities $d_1$ and $d_2$. 

The full generative process of the hierarchical graph structural edge partition (G-HSEPM) model is given by 

\begin{alignat}{2}
&{b}_{ij}^{(t)} &&= \mathbbm{1}({x}_{ij}^{(t)}\geq1),  \\
&{x}_{ij}^{(t)} &&\sim \text{Pois}\left(\sum_{k=1}^{K}r_{k}^{(t)} \phi_{ik} \phi_{jk}\right), \\
&\phi_{ik} &&\sim \text{Gam}(a_0, 1/c_{i}), \\
&r_{k}^{(t)} &&\sim \text{Gam}\left(\sum_{k_2=1}^{K} \pi_{kk_2} r_{k_2}^{(t-1)}, 1/\tau\right),\text{for t = 2,} \cdots \text{, T}, \\
&r_{k}^{(1)} &&\sim \text{Gam}\left(1/K, 1/\tau\right), \\
&\pi_k &&\sim \text{Dir}\left(\alpha_{1k},\cdots,\alpha_{Kk} \right), \\
& \alpha_{k_1k_2} &&=  z_{k_1k_2} \cdot w_{k_1k_2}, \\ 
& w_{k_1k_2} && \sim \text{Gam}(e_0,e_0), \\
& z_{k_1k_2} &&= \mathbbm{1}(\omega_{k_1k_2}\geq1), \\
& \omega_{k_1k_2} &&\sim \text{Pois} \left(\sum_{d_1=1}^{D} \sum_{d_2=1}^{D} m_{k_1d_1} v_{d_1d_2} m_{k_2d_2}\right), \\
& m_{kd} &&\sim \text{Gam}\left(a_k, 1/c_k\right), \\
& c_{k} &&\sim \text{Gam}\left(1, 1\right), \\
& v_{d_1d_2} &&\sim 
\begin{cases}
\text{Gam}(\xi\lambda_{d_{1}}, 1/\beta) & d_1 = d_2 \\
\text{Gam}(\lambda_{d_{1}}\lambda_{d_{2}}, 1/\beta) & d_1 \neq d_2
\end{cases}, \\ 
& \lambda_d &&\sim \text{Gam}\left(\gamma_1/D, 1/c_0\right). \\
\end{alignat}

For the hyperparameters in this framework, we draw them independently from a Gamma distribution, i.e., $a_0, c_i, e_0, c_k, c_o, \beta  \stackrel{i.i.d}{\sim} \text{Gam} (f_0,1/g_0)$. The graphical diagram of this model (G-HSEPM) is illustrated in Figure  \ref{Working_Digram}(right).

\section{Related Work}
Our study focuses on tracking the evolution of dynamic networks in the latent space using a Bayesian framework. In this section, we discuss several previous works that are closely aligned with our approach. The mixed-membership stochastic blockmodel (MMSB) \cite{airoldi2008mixed} is a nonparametric Bayesian extension of the stochastic block model (SBM), allowing each vertex to belong to multiple latent communities. However, MMSB requires the inference of two community indicators for each pair of nodes, regardless of whether there is an edge between them. The hierarchical gamma process edge partition model (HGP-EPM) \cite{zhou2015infinite} was introduced to partition only observed edges. This model associates each observed edge with a latent count via a Bernoulli-Poisson link, and then factorises the latent count, yielding a nonnegative feature matrix that supports an unlimited number of communities. Consequently, each vertex can be affiliated with multiple latent communities, while also being hard-assigned to the single community most strongly represented at its edges. Building on HGP-EPM, Acharya et al. \cite{acharya2015nonparametric} proposed the dynamic gamma process Poisson factorisation for networks (D-NGPPF), which imposes a gamma-Markov chain on community weights, allowing communities to evolve smoothly over time. Another related approach is the dynamic Poisson gamma membership model (DPGM) \cite{yang2018poisson}, which tracks the underlying dynamic network structure by assuming that vertex-community memberships, rather than the community structure itself, are non-static. The probabilistic generative model for overlapping community detection on temporal dual-attributed networks (PGMTAN) \cite{wang2022temporal} also factorises the dynamic node-community membership, while incorporating node attributes to enhance the discovery of significant community structures. Instead of capturing network dynamics using evolving vertex-community memberships, our proposed models assume the underlying community structure is evolving over time, and interacting with each other through a transition kernel, while the vertex-community memberships are static.

\section{Inference}
\label{Inference}
For parameter inference, we introduce a tractable yet efficient Gibbs sampling procedure to obtain model parameter samples from their posterior distributions. The full derivation of the Gibbs sampling algorithms for G-HSEPM and HSEPM are provided in the Appendix \ref{Gibbs_G-HSEPM} and \ref{Gibbs_HSEPM}, respectively.

\section{Experiments}

In this section, we evaluate our model's performance in terms of missing link prediction and community detection. Besides, we demonstrate how the model captures the latent interactions between communities over time. For comparison, we selected three baseline models: the dynamic Poisson gamma membership model(DPGM) \cite{yang2018poisson}, the dynamic gamma process Poisson factorisation for networks (D-NGPPF) \cite{acharya2015nonparametric} and the hierarchical gamma process edge partition model (EPM) \cite{zhou2015infinite}. We used their default parameters settings as released code online. 

We generated one synthetic dataset and then selected four real-world datasets for the experiments. Each dataset was reformulated into a sequence of binary matrices of size $\vert \mathbb{V} \vert\times \vert \mathbb{V} \vert$, where the length of the sequence is time steps $T$. The relationship between two nodes is represented by a binary link, where 1 indicates an association between the nodes and 0 indicates no association. Table \ref{tab:dataset} summarises the number of nodes, edges and time steps of each dataset. More information on each dataset is provided below.

\begin{table}[h]
    \centering
    \setlength{\tabcolsep}{4pt}
    \begin{tabular}{@{}llllll@{}}
    \toprule
         & vdBunt & Synthetic &  Enron & emailEu & DBLP \\
        \midrule
        Nodes   & 32 & 60 & 151  & 274  & 933   \\
        Edges   & 308 & 2,520 & 2,799 &  11,266  & 16,792  \\
        Time   & 7 & 6 & 20  & 18  & 10  \\
        \bottomrule
    \end{tabular}
    \caption{The sufficient statistics of the datasets.}
    \label{tab:dataset}
\end{table}

\noindent \textbf{Students datasets in van de Bunt (vdBunt)}. This dataset was collected by Gerhard van de Bunt \cite{van1999friendship}, recording the friendship among 32 previously unacquainted university freshmen across 7-time steps, where we reshaped it as a $7 \times 32 \times 32$ matrix.


\noindent \textbf{Synthetic data}. We generated a dynamic network with $N = 60$ vertices, evolving over $T = 6$ time steps. At time steps 1, 3 and 5, there are a large community and four smaller communities. At time steps 2, 4 and 6, the second smaller community will interact with the first large community. We also generated the random addition and removal edges in each community to simulate the real-world vertex-community relationships. Finally, we got a $6 \times 60 \times 60$ dynamic network.

\noindent \textbf{Enron email communication (Enron)}. This dataset recorded around half a million email communications among 2,359 people for 28 months\cite{tang2008networks}. We filtered out individuals whose email records fewer than 5 snapshots and selected the 20 consecutive months with the highest email activity. The resulting data was reshaped as a $20\times151\times151$ binary symmetric matrix, indicating the presence or absence of email communication between two individuals at each time step.

\noindent \textbf{Email-Eu-core temporal network (emailEu)}. This dataset was generated from a large European research institution, recording the incoming and outgoing email of the research institution from October 2003 to May 2005 (18 months) \cite{leskovec2007graph}. Similar to the Enron dataset, we selected individuals with at least 50 records as either email senders or receivers. Finally, we maintained $N = 274$ individuals and constructed a $18 \times 274 \times 274$ matrix.

\noindent \textbf{DBLP conference abstracts (DBLP)}. This dataset recorded the co-authorships among 347,013 authors in the DBLP database\cite{tang2008networks}. We selected the authors who have had consecutive publication activities over the last 10 years and then filtered out authors who have less collaboration with others. Finally, there were 933 authors left and we obtained a $10 \times 933 \times 933$ matrix.

\subsection{Missing link prediction}
We compared the missing link prediction performance of our models against baseline models using four real-world datasets: \emph{vdBunt}, \emph{Enron}, \emph{EuEmail}, and \emph{DBLP} datasets. For these datasets, we initialised the number of communities $K$ to 10, 50, 100 and 100, respectively. Meanwhile, we set the number of hierarchical communities $D$ to 5, 30, 50 and 50, respectively. For each dataset, we randomly selected 70\% of the data for training and masked the remaining 30\% for testing. We conducted 10 independent runs for each model, with each run including 3000 iterations of Gibbs sampling. The first 2000 iterations were for burn-in and the remaining 1000 iterations were used for collection. Our evaluation metrics included accuracy \cite{fisher1936use}, F1-score \cite{powers2020evaluation}, Area Under the Receiver Operating Characteristic Curve (AUC-ROC), and Area Under the Precision-Recall Curve (AUC-PR) \cite{davis2006relationship}, which can comprehensively evaluate the model's performance.

\begin{figure}[hbt!]
    \centering
    \begin{subfigure}{0.55\columnwidth}
        \includegraphics[width=\columnwidth]{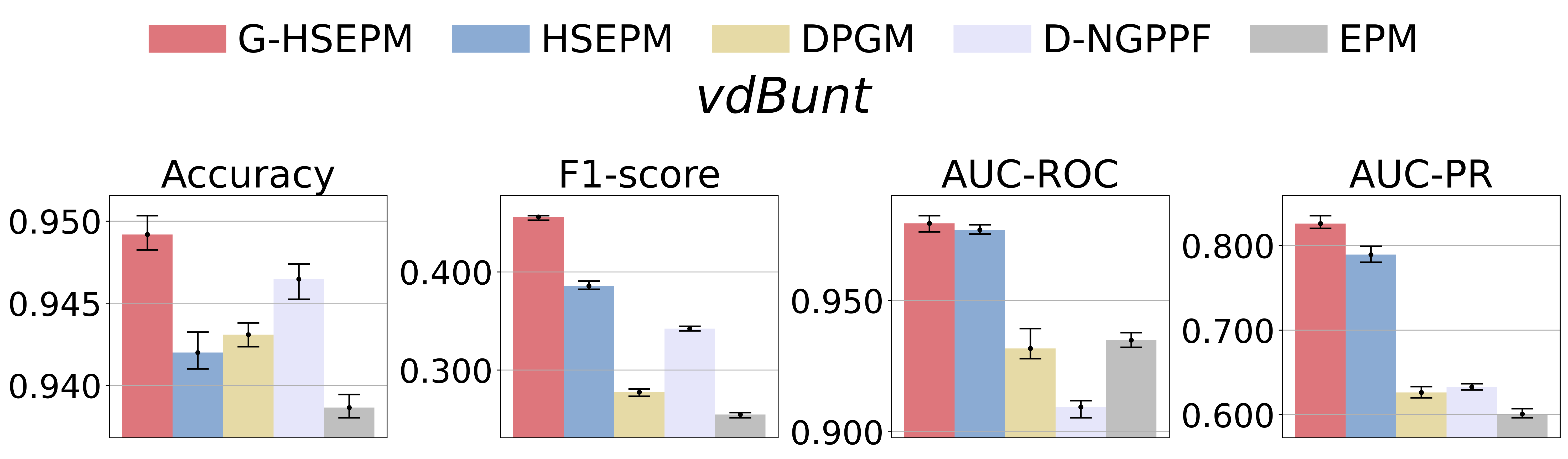}
    \end{subfigure}
    \vspace{0.2cm}
    \begin{subfigure}{0.55\columnwidth}
        \includegraphics[width=\columnwidth]{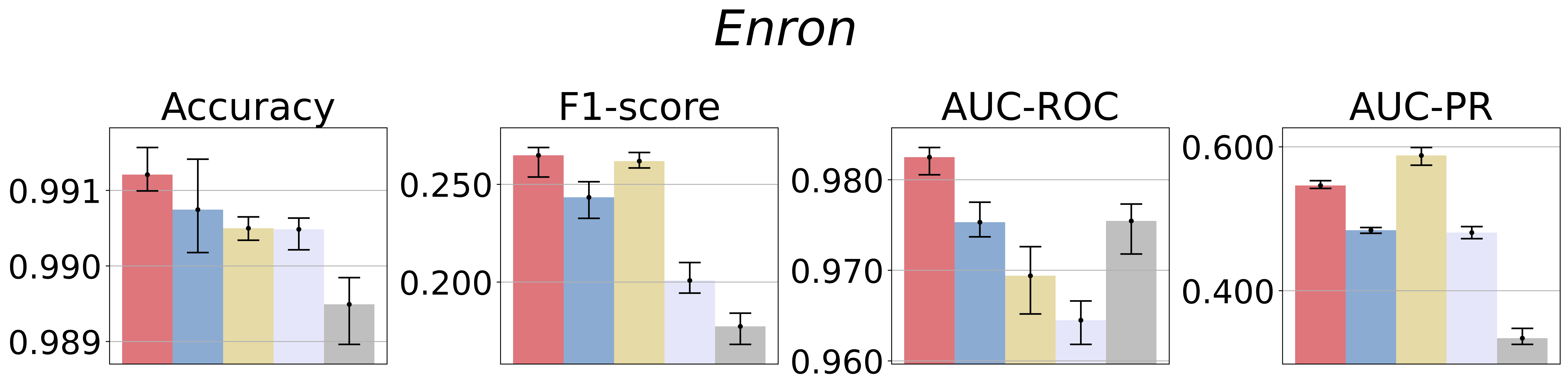}
    \end{subfigure}
    \vspace{0.2cm}
    \begin{subfigure}{0.55\columnwidth}
        \includegraphics[width=\textwidth]{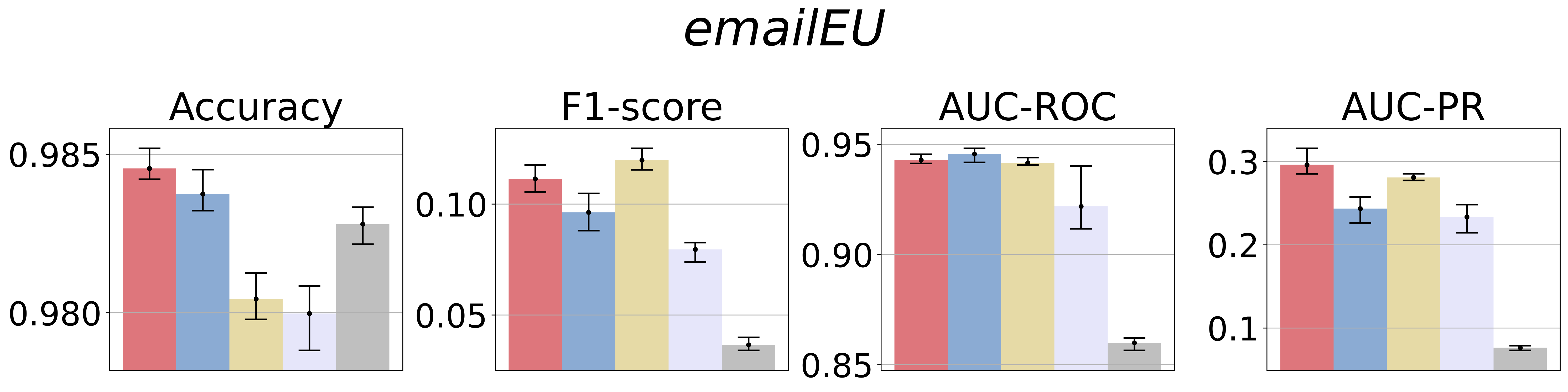}
    \end{subfigure}
    \vspace{0.2cm}
    \begin{subfigure}{0.55\columnwidth}
        \includegraphics[width=\textwidth]{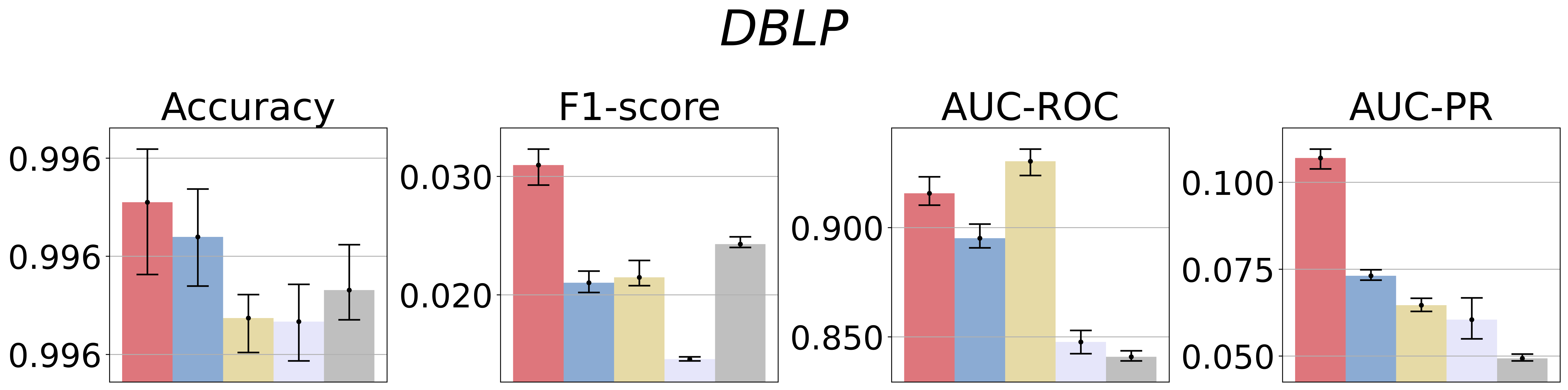}
    \end{subfigure}
    \caption{The comparisons of the model performance in terms of missing links prediction.}
    \label{evaluation}
\end{figure}

Figure \ref{evaluation} compares the performance of each model in different datasets. Overall, our models (G-HSEPM and HSEPM) outperform the others due to several key structural advantages. First, the Markovian framework of G-HSEPM and HSEPM allows them to consider both forward and backward information at each time step, leading to more accurate predictions compared to the static nature of EPM. Second, the transition kernel in G-HSEPM and HSEPM accounts for community interactions when determining community weights, rather than relying solely on the previous time step weights as D-NGPPF. Third, unlike DPGM, which tracks network dynamics via vertex-community memberships and requires the costly update of a $T \times N \times K$ matrix per iteration, our models efficiently capture network dynamics by updating a more compact $T \times K$ community weights matrix in each iteration, significantly reducing computation time while maintaining modelling performance. Table \ref{tab:computation time} compares the average per-iteration computation time of each model (all implemented in Python). Lastly, while G-HSEPM requires more computation time than HSEPM due to its more complex structure, the hierarchical graph structure imposed on the transition kernel leads to more accurate parameter estimation and a more interpretable model of G-HSEPM. Further details on the interpretability of G-HSEPM are provided in the following sections. 

\begin{table}[ht]
    \centering
    \begin{tabular}{llllll}
    \toprule
         & vdBunt & Enron & emailEu  & DBLP \\
    \midrule
    G-HSEPM   & 0.206 & 2.068 & 8.013  & 17.731  \\
    HSEPM   & \textbf{0.165} & \textbf{1.777} & \textbf{6.316}  & \textbf{12.852}  \\
    DPGM   & 0.248 & 2.185 & 11.158  & 20.405 \\
    D-NGPPF  & \textbf{0.132} & \textbf{1.523}  & 9.177  & \textbf{10.179} \\
    HGP-EPM  & 0.268 & 4.481  & 13.017 & 24.084 \\
    \bottomrule  
    \end{tabular}
    \caption{The comparison of per-iteration computation time in seconds.}
\label{tab:computation time}
\end{table}

\subsection{Dynamic community detection}
\label{Dynamic community detection}
\begin{figure*}[h!]
    \centering
    \begin{subfigure}[b]{0.73\columnwidth}
        \includegraphics[width=\columnwidth]{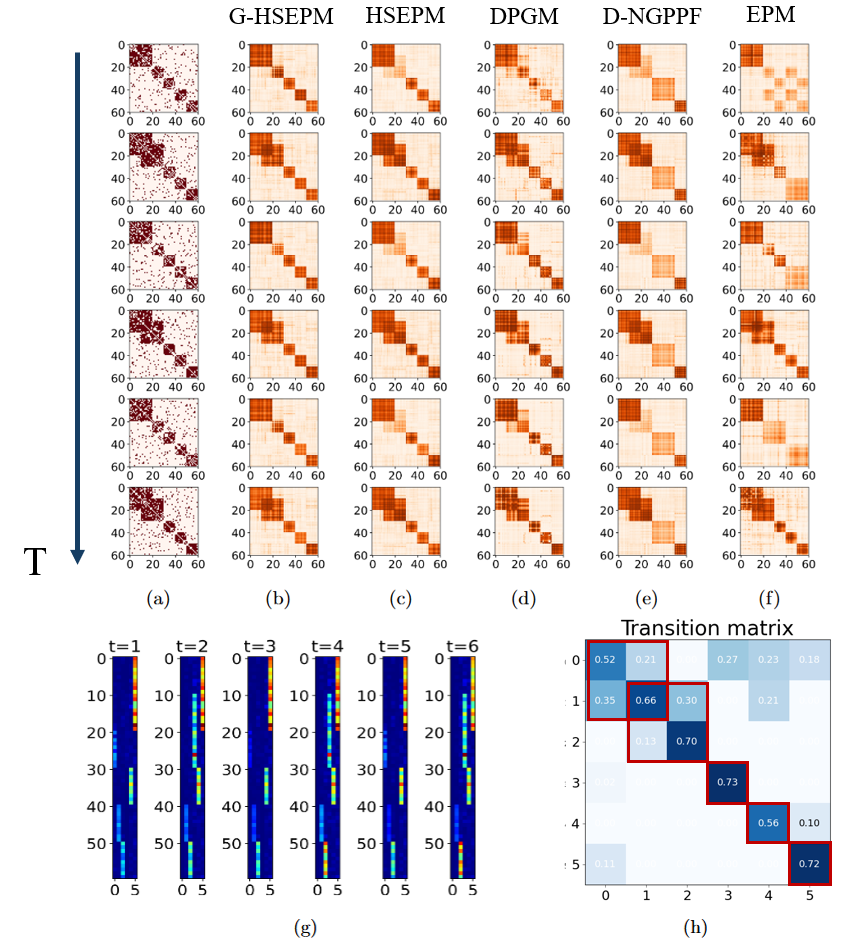}
    \end{subfigure}
    \caption{Dynamic community detection on the synthetic dataset. (a) is the generated dynamic network, whose community structure is evolving with time ordered from top to bottom. (b), (c), (d), (e) and (f) are link probability inferred by G-HSEPM, HSEPM, DPGM, D-NGPPF and EPM, respectively. (g) shows how each vertex is allocated to communities of G-HSEPM at each time. (h) is the transition matrix of G-HSEPM, showing the interactions between communities.}
    \label{Synthetic_data}
\end{figure*}
We evaluated the community detection capabilities of our models in comparison with other baseline models using a synthetic dataset. All models were initialized with $K = 6$ and $D = 6$. Plot (a) of Figure \ref{Synthetic_data} illustrates the structure of synthetic data. Plot (b)-(f) show the link probabilities inferred by G-HSEPM, HSEPM, DPGM, D-NGPPF and EPM, respectively, where we observed that G-HSEPM more accurately captures the community structure at each time step compared to the other models. Plot (g) displays the temporal evolution of vertex-community membership $r_{k}^{(t)} \phi_{ik}$ in G-HSEPM, where we observed that vertex affiliations to communities shift as the community structure evolves. Specifically, when the second community attempts to merge with the first community, the vertices in the overlapping region show an increased probability of belonging to both two communities. Plot (h) represents the transition matrix, which shows the probability of transitioning between communities. A lower transition probability indicates fewer interactions between two communities, and vice versa. In this synthetic dataset, the second community occasionally interacts with the first community, leading to a higher overlap probability between these two communities in the transition matrix. In contrast, the other three communities do not interact with any others, resulting in a higher probability of self-transitions. These observations demonstrate the ability of G-HSEPM to effectively interpret the latent structure of the data.

\begin{figure}[hbt!]
    \centering
    \begin{subfigure}[b]{0.7\columnwidth}
        \includegraphics[width=\columnwidth]{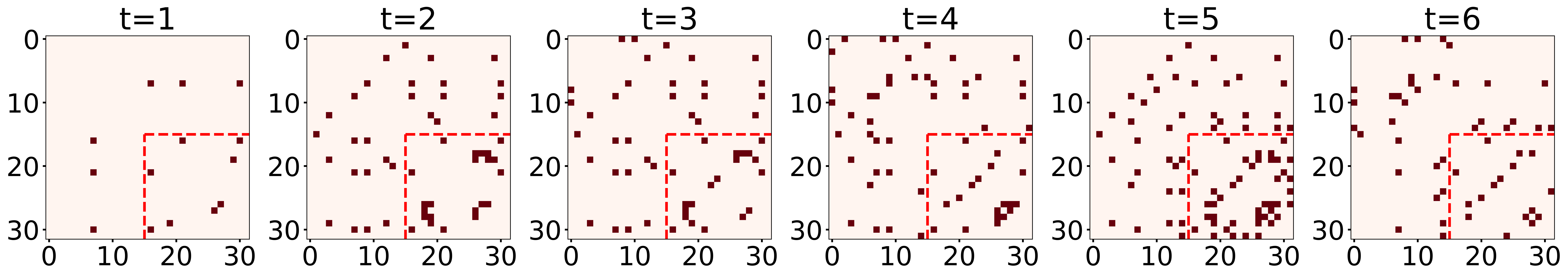}
        \caption{}
    \end{subfigure}
    \vfill
    \begin{subfigure}[b]{0.7\columnwidth}
        \includegraphics[width=\columnwidth]{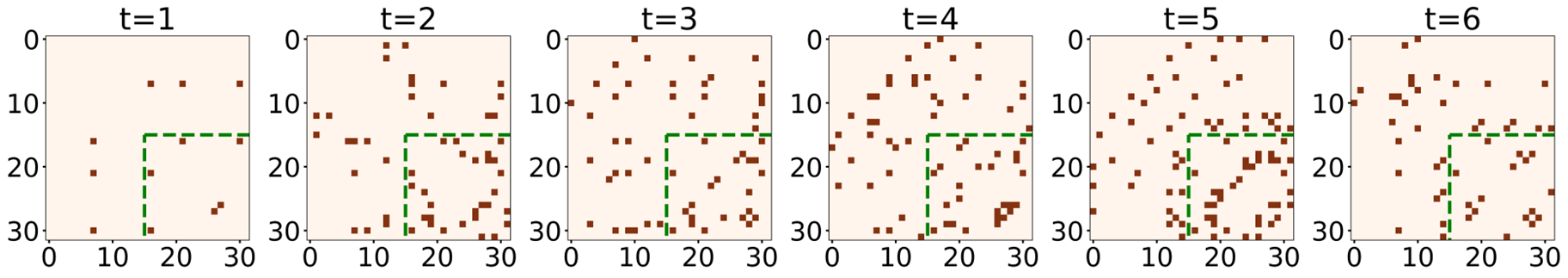}
        \caption{}
    \end{subfigure}
    \vfill
    \begin{subfigure}[b]{0.7\columnwidth}
        \includegraphics[width=\columnwidth]{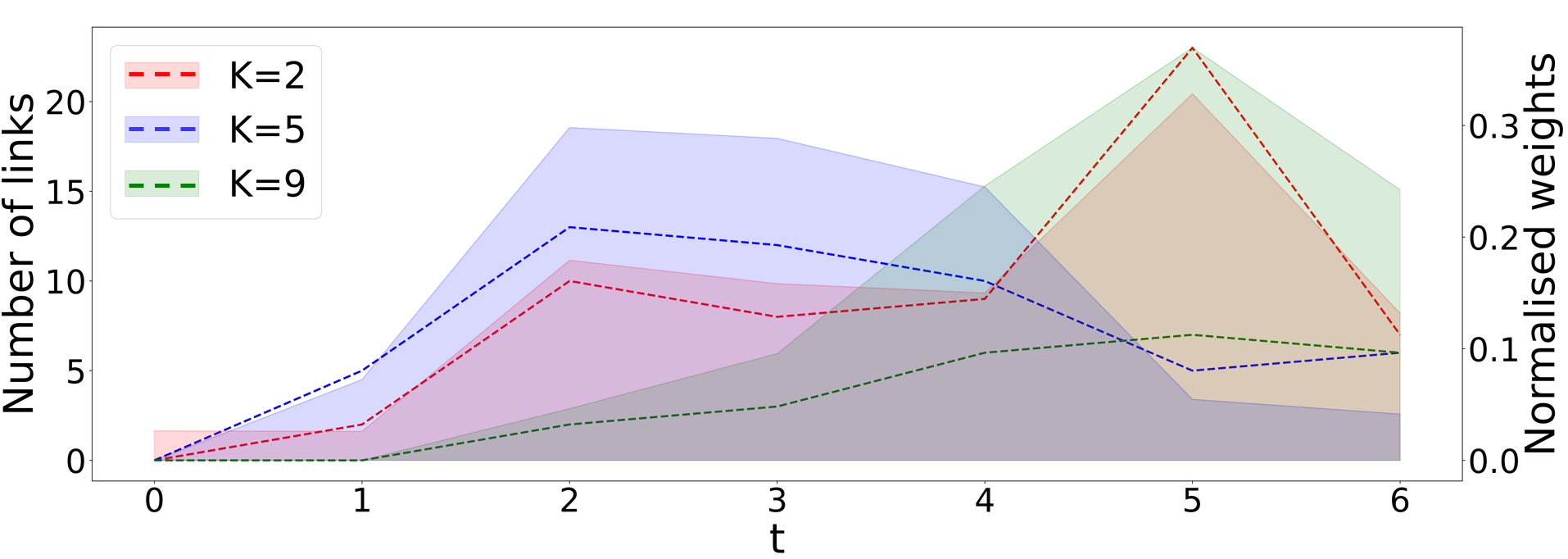}
        \caption{}
    \end{subfigure}   
    \caption{Dynamic community detection on vdBunt dataset. (a) shows the true networks. (b) depicts the networks inferred by our G-HSEPM model. We can find the inferred graph structure is aligned with the true graph. (c) illustrates how the normalised weights $r_{k}^{(t)} / (\sum_{t}r_{k}^{(t)})$ of the top 3 communities and the number of links between their corresponding major vertices change over time. The red, blue and green colours represent the top communities 2, 5, and 9, respectively. Dotted lines represent the number of links, and the colour-filled areas represent the normalised community weights.}
    \label{Stu_CommunityDetection}
\end{figure}

We also performed the community detection on the real-world dataset, as depicted in Figure \ref{Stu_CommunityDetection}. 
Plot (a) shows the graph structure of the vdBunt dataset. In the lower-left corner, where we had highlighted, we identified a specific structure that evolves over time. Plot (b) demonstrates that our model effectively captures this evolving structure. In plot (c), we focused on the top 3 communities with the highest weights and identified the top 10 major nodes in each corresponding community. We observed that the temporal variation in community weights closely aligns with the changes in the number of links between their corresponding vertices. This indicates that the community evolution is correlated with the interactions between vertices as we expected. This pattern also appears in other datasets, which are provided in the Appendix \ref{Additional Results}.

\subsection{Exploratory analysis}
\label{Exploratory analysis}

\begin{figure}[hbt!]
    \centering
    \begin{subfigure}[b]{0.25\columnwidth}
        \includegraphics[width=\columnwidth]{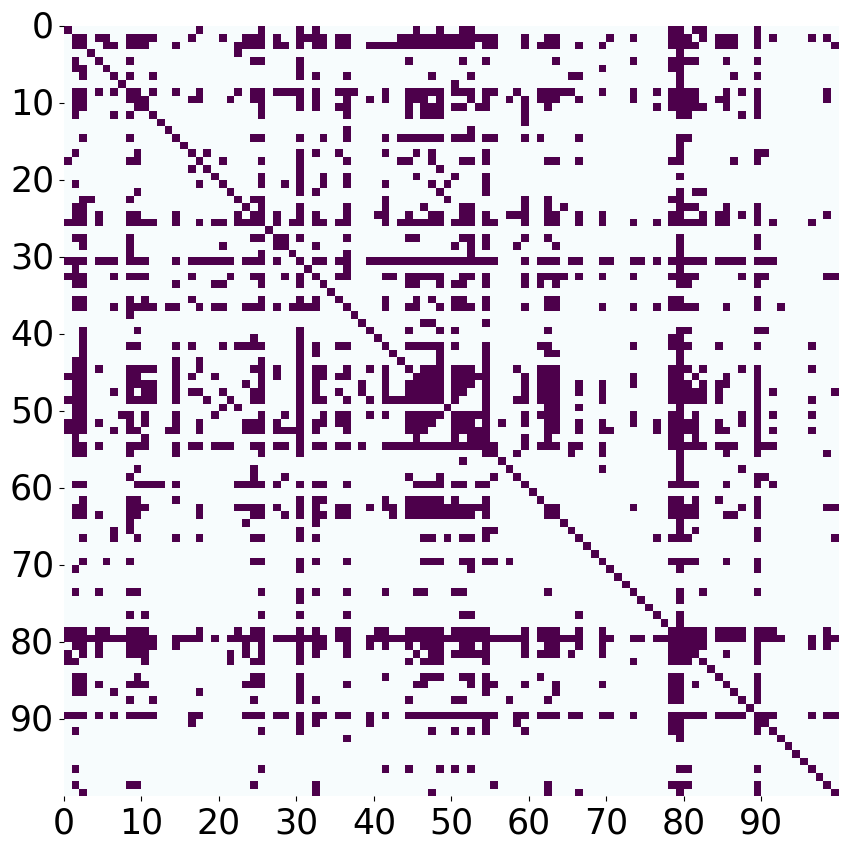}
        \caption{The connections between communities $Z_{k_1k_2}$}
    \end{subfigure}
    \hspace{1.5cm}   
    \begin{subfigure}[b]{0.25\columnwidth}
        \includegraphics[width=\columnwidth]{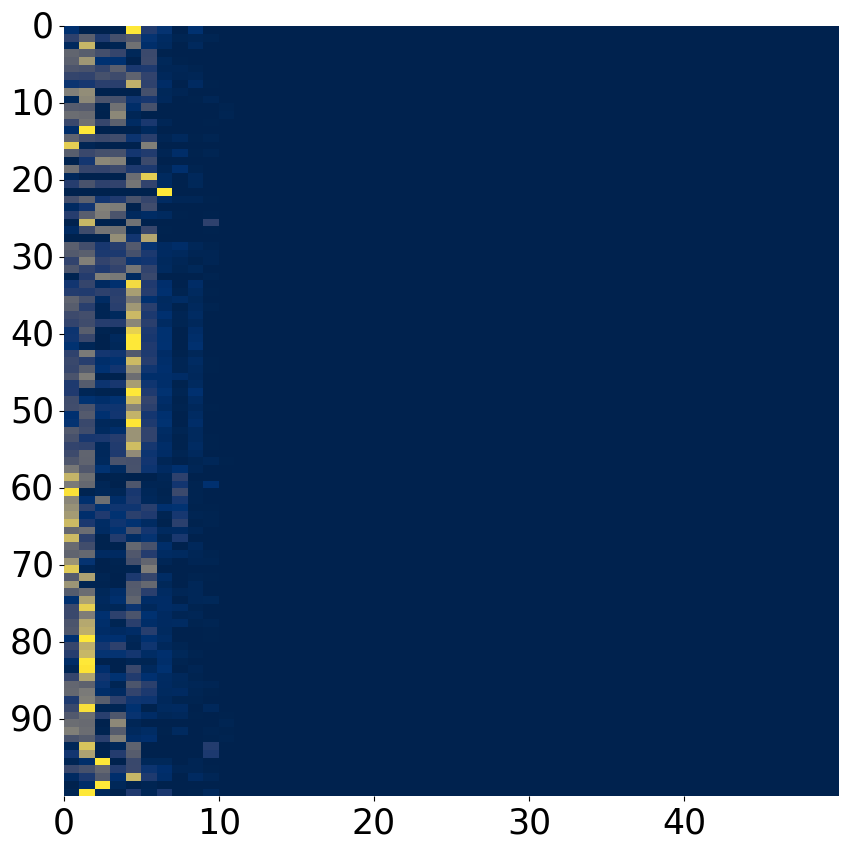}
        \caption{The probabilities of 100 communities affiliation to 50 latent hierarchical communities $m_{kd}$.}
    \end{subfigure}
    \caption{The latent graph structure of DBLP dataset inferred by G-HSEPM.}      
    \label{DBLP_LatentStructure}
\end{figure}

As discussed in the previous sections, we constructed a graph structure within G-HSEPM to describe the relationships between communities, resulting in a more flexible and interpretable model. In this section, we analysed the latent communities and their hierarchical structure of the DBLP dataset, as shown in Figure \ref{DBLP_LatentStructure}. 
Plot (a) presents the relationships between community $k_1$ and community $k_2$, which is denoted by $Z_{k_1k_2}$. However, it is still a challenge to analyse the interactions among a large number of communities. So we aggregated the 100 communities ($K = 100$) into 50 hierarchical communities ($D = 50$). Interestingly, as shown in plot (b), the model naturally assigned these communities to only 7 hierarchical communities. This suggests that communities within the same hierarchical group exhibit strong interconnections.

\begin{figure}[h!]
    \centering
    \begin{subfigure}[b]{0.2\columnwidth}
        \includegraphics[width=\columnwidth]{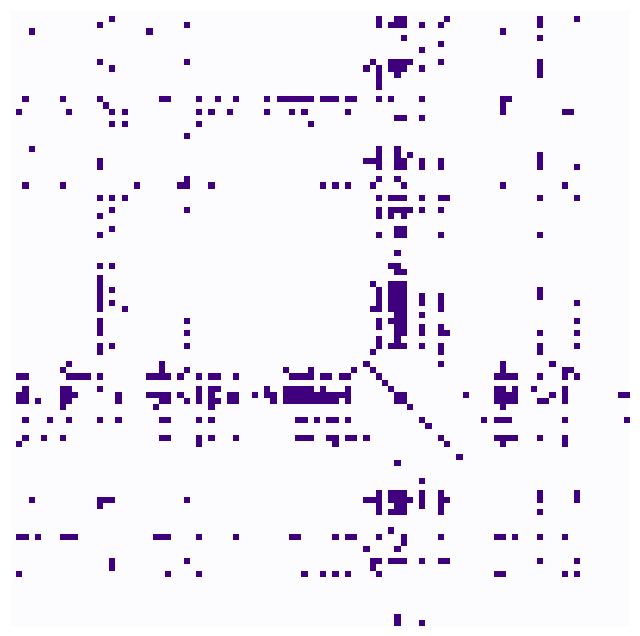}
    \end{subfigure} 
    \hspace{2.0cm}
    \begin{subfigure}[b]{0.2\columnwidth}
        \includegraphics[width=\columnwidth]{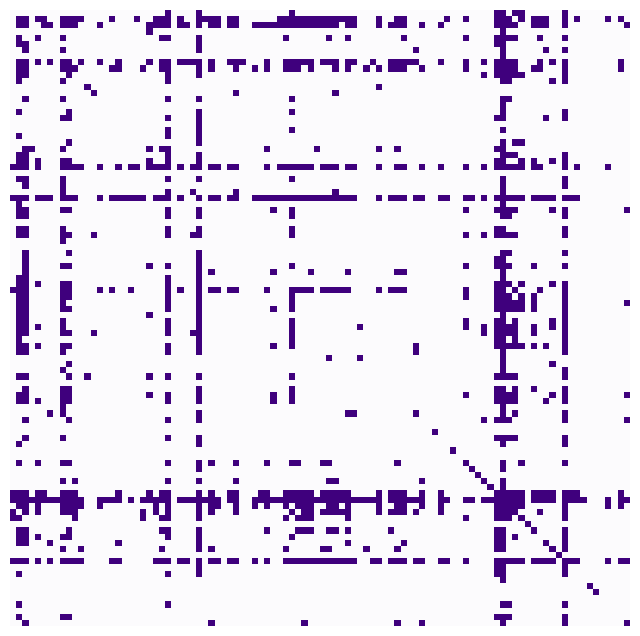}
    \end{subfigure}
    \hspace{2.3cm}
    \begin{subfigure}[b]{0.2\columnwidth}
        \includegraphics[width=\columnwidth]{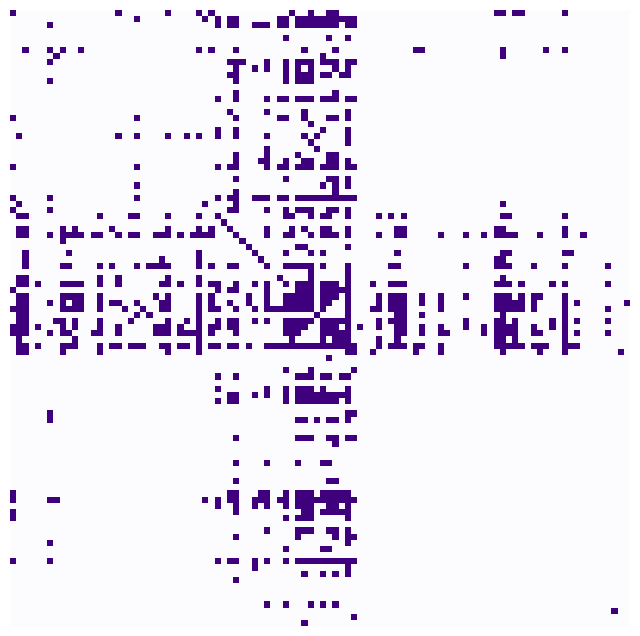}
    \end{subfigure}
    \vfill
    \begin{subfigure}[b]{0.32\columnwidth}
        \includegraphics[width=\columnwidth]{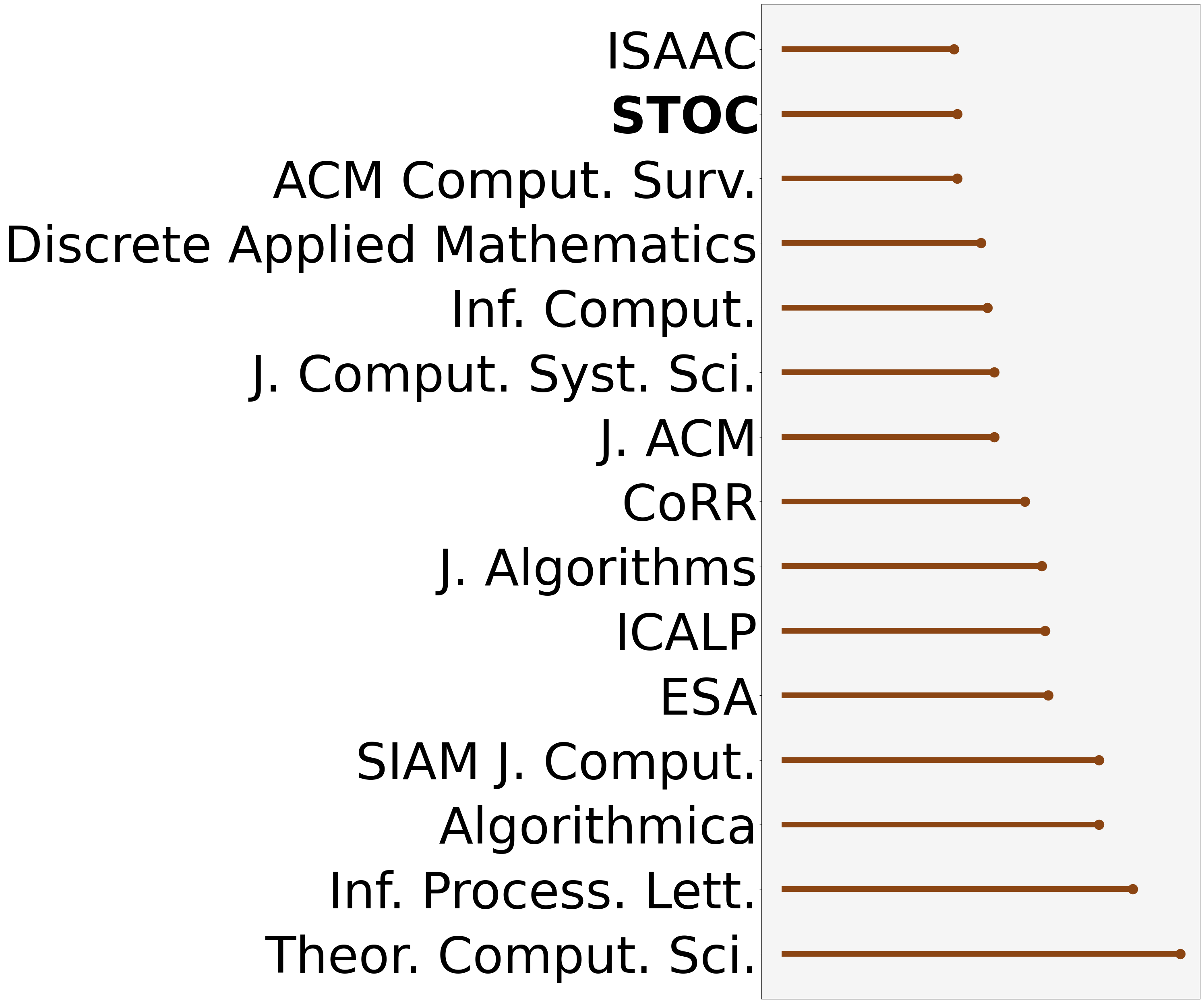}
        \caption{}
    \end{subfigure}
    \hspace{0.1cm}
    \begin{subfigure}[b]{0.32\columnwidth}
        \includegraphics[width=\columnwidth]{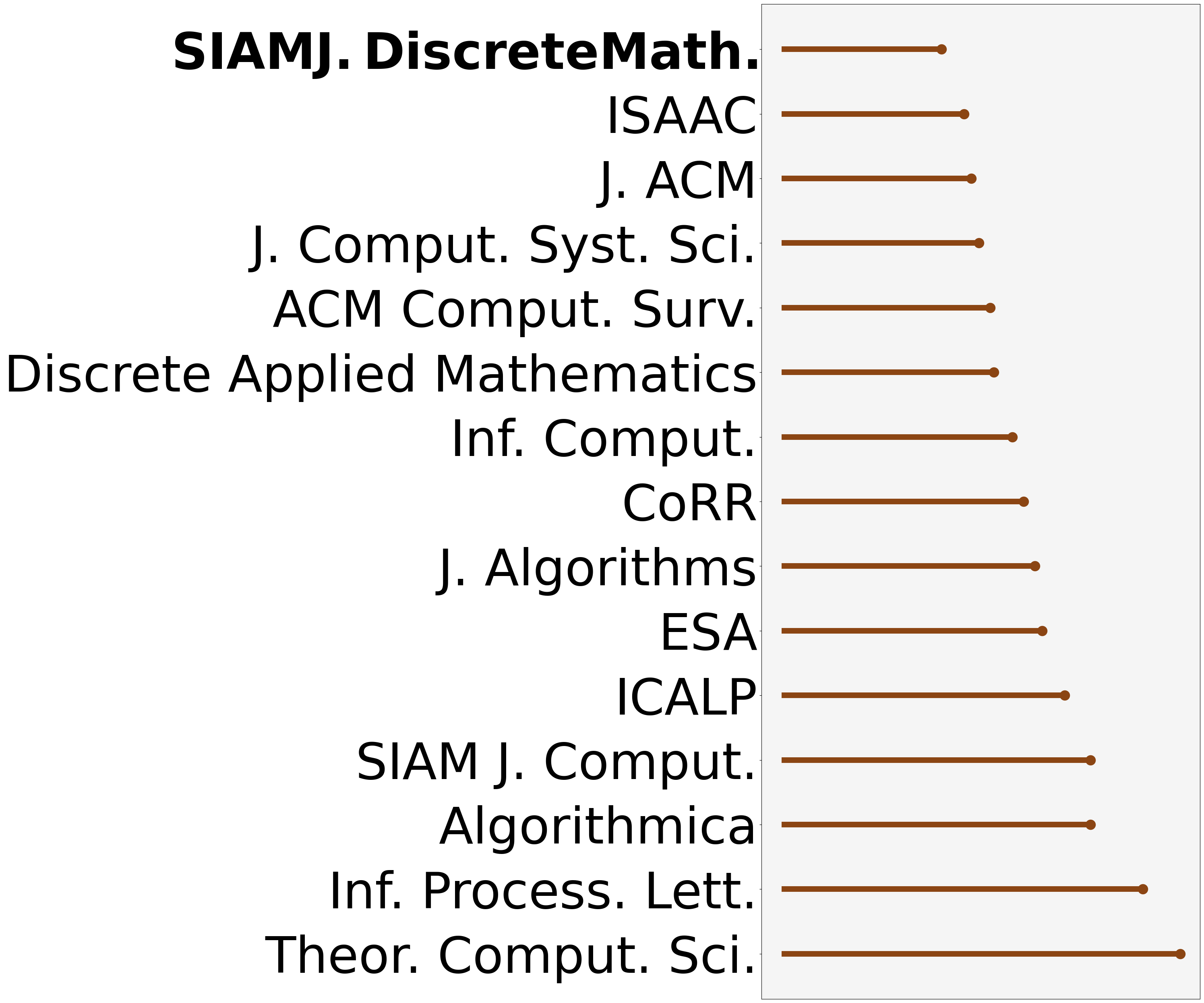}
        \caption{}
    \end{subfigure}
    \hspace{0.1cm}
    \begin{subfigure}[b]{0.32\columnwidth}
        \includegraphics[width=\columnwidth]{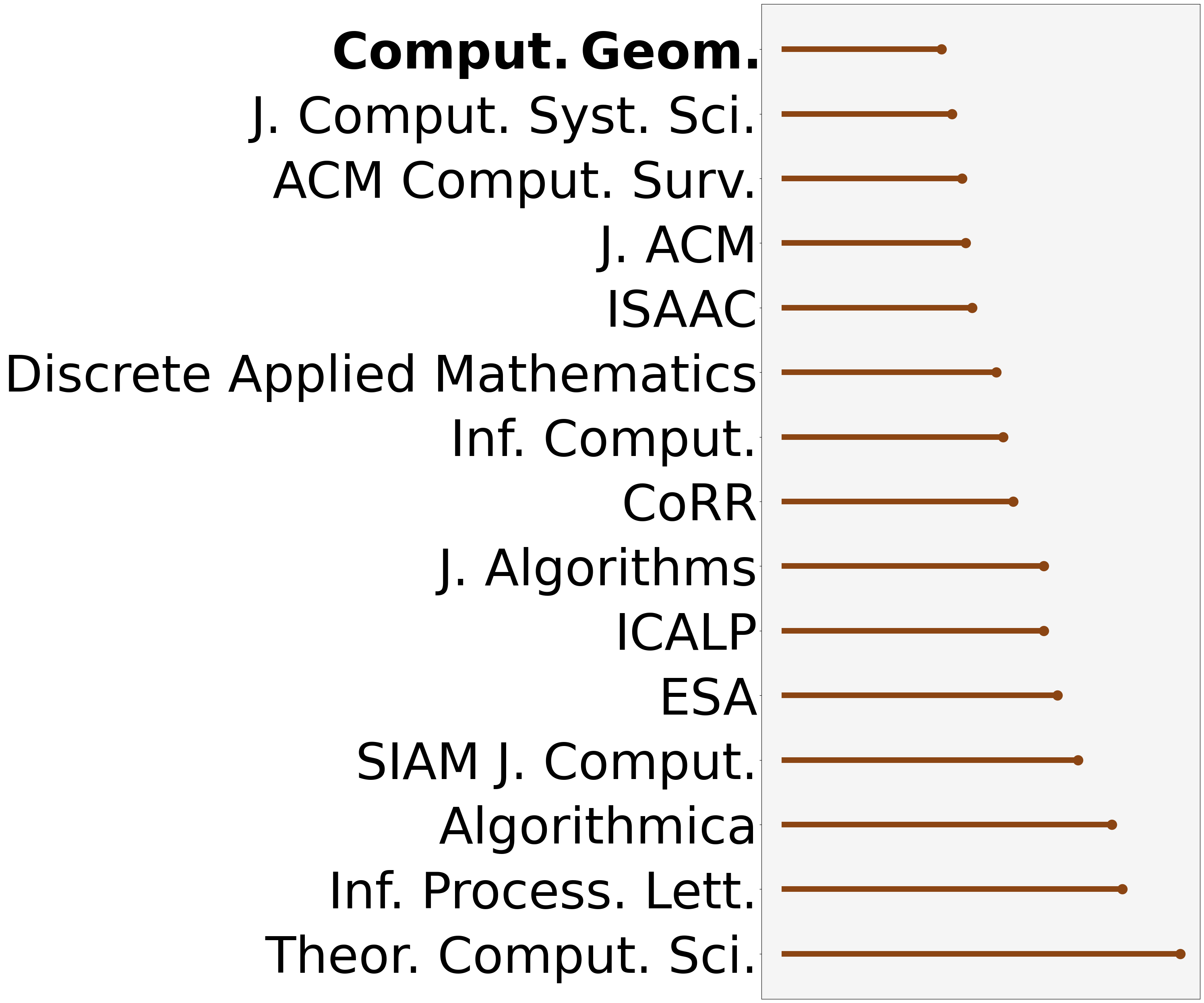}
        \caption{}
    \end{subfigure}

    \caption{Plots (a), (b), and (c) represent the top 3 latent hierarchical communities with the highest number of communities. The plots on the top show the relationship between communities within each latent hierarchical community. The plots on the bottom illustrate the journals and conferences where the corresponding communities most frequently published over 10 years.}      
    \label{DBLP_Journal}
\end{figure}
To further explore the characteristics of these highly connected communities, we selected the top 3 hierarchical communities containing the most communities, and identified their main research fields by listing the top 15 most frequently published journals and conferences, as shown in Figure \ref{DBLP_Journal}. Details of the remaining hierarchical communities and their corresponding publication venues are provided in the Appendix \ref{Additional Results}.

In Figure \ref{DBLP_Journal}, the top heatmaps display the relationships among communities within each hierarchical community, while the bottom plots present their most frequently published journals and conferences. We found that these 3 hierarchical communities share 14 common publication venues in computer science and mathematics. However, their distinguishing factor lies in their specific focus: communities in the first hierarchical community (plot (a)) frequently publish in the \emph{Symposium on Theory of Computing (STOC)}, which is oriented toward theoretical computer science; the second hierarchical community (plot (b)) frequently publishes in the \emph{SIAM Journal on Discrete Mathematics (SIAM J. Discrete Math.)}, focusing on discrete mathematics; and the third hierarchical community (plot (c)) frequently appears in \emph{Computational Geometry: Theory and Applications (Comput.Geom.)}, emphasising computational geometry. These observations suggest that, while the communities share some common fields, our model can still effectively distinguish them based on their different research areas.

\section{Conclusion}
This paper has developed a novel dynamic network model, to capture how the underlying communities evolve and interact with each other over time. In particular, a sparse-graph-structured-transition kernel is dedicated to modelling the hierarchical interactions behind those estimated latent communities. Tractable-yet-efficient Gibbs sampling algorithms are designed to perform posterior inference for the studied methods. The final experiments show the excellent performance of the novel model, compared to prior related works, in terms of link prediction, community detection and model explainability. In future research, we will investigate how to capture dynamic networks, in which the number of vertices can also change over time, and the network edges might be sampled at an irregularly-spaced time axis. In addition, another interesting direction is to consider modeling how the dynamic network snapshot observed over regularly-spaced time steps, and its edges created continuously~\cite{DBLP:conf/uai/YangK20,DBLP:conf/aaai/YangZ24,DBLP:conf/sdm/YangZ23}, interact or influence with each other.  


\bibliographystyle{unsrt}  

\appendix
\section{Gibbs sampling for G-HSEPM}
\label{Gibbs_G-HSEPM}
The sampling algorithm depends on three data augmentation and marginalisation techniques to derive a closed-form conditional posterior.

\noindent \textbf{\emph{Lemma 1.}} If $x.=\sum_{n=1}^N x_n$, where $x_n \sim \text{Pois}(\lambda_n)$ are independently drawn from a  Poisson distribution, then $(x_1,\cdots,x_N) \sim \text{Mult} (x.,(\frac{\lambda_1}{\sum_{n=1}^N \lambda_n}, \cdots ,\frac{\lambda_N}{\sum_{n=1}^N \lambda_n}))$ and $x. \sim \text{Pois}(\sum_{n=1}^N \lambda_n)$ \cite{dunson2005bayesian}.

\noindent \textbf{\emph{Lemma 2.}} If $x \sim \text{Pois}(k\lambda)$, where $k$ is a constant and  $\lambda \sim \text{Gam}(s,r)$, then $x$ can also be sampled from a negative binomial (NB) distribution as $x \sim \text{NB}(s,\frac{k}{r+k})$. Equivalently, $x \sim \text{NB} (s,1-\exp{(-\zeta)})$, where $\zeta = \ln{(1+\frac{k}{r})}$ \cite{schein2016poisson}

\noindent \textbf{\emph{Lemma 3.}} If $x \sim \text{NB}(s,1-\exp{(-\zeta)})$, and $l$ is sampled from a Chinese restaurant table distribution $l \sim \text{CRT}(x,s)$, then, $x \sim \text{SumLog}(l, 1-\exp{(-\zeta)})$ and $l \sim \text{Pois}(s\zeta)$.  \cite{zhou2013negative}.

Here we present the Gibbs sampling for the G-HSEPM model. 

\noindent \textbf{Sampling latent counts $x_{ij}^{(t)}$.} We sample the latent count for each observed edge $b_{ij}^{(t)}$ at each timestamp as

\begin{equation}
    (x_{ij}^{(t)}|-) \sim b_{ij}^{(t)}\text{Po}_{+}(\sum_{k=1}^{K}r_k^{(t)}\phi_{ik}\phi_{jk}).
    \label{xij}
\end{equation}

Since $x_{ij}^{(t)} = \sum_{k=1}^{K} x_{ijk}^{(t)}$, using the Poisson additive property, $x_{ijk}^{(t)} \sim \text{Pois} (r_k^{(t)}\phi_{ik}\phi_{jk})$. Via the Poisson-multinomial equivalence in \emph{Lemma 1}, the latent count $x_{ijk}^{(t)}$ can be sampled as

\begin{equation}
    (x_{ijk}^{(t)}|-) \sim \text{Mult} (x_{ij}^{(t)},(\frac{r_k^{(t)}\phi_{ik}\phi_{jk}}{\sum_{k=1}^K r_k^{(t)}\phi_{ik}\phi_{jk}})).
    \label{xijk}
\end{equation}

\noindent \textbf{Sampling membership of nodes and communities $\phi_{ik}$.} Via the gamma-Poisson conjugacy, we obtain the posterior distribution of $\phi_{ik}$ as

\begin{equation}
    (\phi_{ik}|-) \sim \text{Gam} (a_0 + \sum_{t=1}^{T} \sum_{j\neq i}^{N} x_{ijk}^{(t)}, 1/(c_i + \sum_{t=1}^{T} \sum_{j\neq i}^{N} r_{k}^{(t)}\phi_{jk})).
    \label{phi}
\end{equation}

\noindent \textbf{Sampling hyperparameter $c_i$.} Via the gamma-gamma conjugacy, we sample $c_i$ as 

\begin{equation}
    (c_i|-) \sim \text{Gam} (f_0 + Ka_0, 1/(g_0 + \sum_{k=1}^{K}\phi_{ik}))
    \label{ci}
\end{equation}

\noindent \textbf{Sampling community weights $r_{k}^{(t)}$.} Since the community weights $r_{k}^{(t)}$ evolves in a Markovian construction, the backward and forward information need to be incorporated into updates. 

For $t=T$, $x_{..k}^{(T)} = \sum_{i}^{N-1} \sum_{j=(i+1)}^{N} x_{ijk}^{(T)}$, $x_{..k}^{(T)} \sim \text{Pois} (r_k^{(t)}s_k)$, where $s_k = \sum_{i}^{N-1} \sum_{j=(i+1)}^{N} \phi_{ik}\phi_{jk} $. Via gamma-Poisson conjugacy

\begin{equation}
    (r_k^{(T)}|-) \sim \text{Gam} (x_{..k}^{(T)} + \sum_{k_2=1}^{K} \pi_{kk_2} r_{k_2}^{(T-1)}, 1/(\tau + s_k)).
    \label{rkT}
\end{equation}

For $t=T-1$, to get the forward information, we marginalise out $r_k^{(T)}$

\begin{equation}
    x_{..k}^{(T)} \sim \text{NB} (\sum_{k_2=1}^{K} \pi_{kk_2} r_{k_2}^{(T-1)}).
\end{equation}
According to \emph{Lemma 3}, the negative binomial distribution can be augmented with an auxiliary variable 

\begin{equation}
    l_{.k}^{(T)} \sim \text{CRT} (x_{..k}^{(T)}, \sum_{k_2=1}^{K} \pi_{kk_2} r_{k_2}^{(T-1)}).
    \label{lTk}
\end{equation}
We can then re-express the joint distribution over $x_{..k}^{(T)}$ and $l_{k}^{(T)}$ as 
\begin{equation}
    x_{..k}^{(T)} \sim \text{SumLog}(l_{k}^{(T)}, \rho_k^{(T)}),
\end{equation}

\begin{equation}
    l_{k}^{(T)} \sim \text{Pois}\left(-\sum_{k_2=1}^{K} \pi_{kk_2} r_{k_2}^{(T-1)} \ln(1-\rho_k^{(T)})\right).
\end{equation}
where $\rho_k^{(T)} = \frac{s_k}{\tau + s_k}$. Since $l_k^{(T)} = l_{k.}^{(T)} = \sum_{k_2=1}^{K} l_{kk_{2}}^{(T)}$ and $l_{.k}^{(T)} = \sum_{k_1=1}^{K}l_{k_{1}k}^{(T)}$, via \emph{Lemma 1}, we can express the distribution $l_{.k}^{(T)}$ as $l_{.k}^{(T)} \sim \text{Pois}(-r_k^{(T-1)}\ln(1-\rho_k^{(T)}))$.

Given $x_{..k}^{(T-1)} \sim \text{Pois} (r_k^{(T-1)}s_k)$, via the Poisson additive property, we have 

\begin{equation}
    x_{..k}^{(T-1)} + l_{.k}^{(T)} \sim \text{Pois} (r_k^{(T-1)} (s_k - \ln(1-\rho_k^{(T)}))).
\end{equation}
This equation summarises the backward and forward information at time $T-1$. Combing with the gamma prior placed on $r_k^{(T-1)}$, we obtain its conditional distribution via the gamma-Poisson conjugacy as

\begin{equation}
\begin{aligned}
    (r_k^{(T-1)}|-) &\sim \text{Gam} \left( x_{..k}^{(T-1)} + l_{.k}^{(T)} + \sum_{k_{2}}^{K}\pi_{kk_{2}}r_{k_{2}}^{(T-2)}, \right. \\
    &\quad \left. 1/(\tau + s_k - \ln(1-\rho_k^{(T)})) \right).
\end{aligned}
\end{equation}

For $t=T-2,\cdots,2$, we introduce
\begin{equation}
    l_{.k}^{(t)}\sim\text{CRT} (x_{..k}^{(t)} + l_{.k}^{(t+1)}, \sum_{k_2=1}^{K} \pi_{kk_2} r_{k_2}^{(t-1)}),
    \label{ltk}
\end{equation}
and do a similar augmentation and sampling trick as what we do for $t = T$.

\begin{equation}
\begin{aligned}
    (r_k^{(t)}|-) &\sim \text{Gam} \left( x_{..k}^{(t)} + l_{.k}^{(t+1)} + \sum_{k_{2}}^{K}\pi_{kk_{2}}r_{k_{2}}^{(t-1))}, \right. \\
    &\quad \left. 1/(\tau + s_k - \ln(1-\rho_k^{(t+1)})) \right),
    \label{rkt}
\end{aligned}
\end{equation}
where 
\begin{equation}
    \rho_k^{(t+1)}=\frac{s_k-\ln{(1-\rho_k^{(t+2)})}}{\tau+s_k-\ln{(1-\rho_k^{(t+2)})}}. 
    \label{rho}
\end{equation}

For $t=1$, similarly, augment 

\begin{equation}
    l_{k}^{(1)} \sim \text{CRT} (x_{..k}^{(1)}+l_{.k}^{(2)}, 1/K),
    \label{l1k}
\end{equation}
then sample

\begin{equation}
\begin{aligned}
    (r_k^{(1)}|-) &\sim \text{Gam} \left( x_{..k}^{(1)} + l_{.k}^{(2)} + 1/K, \right. \\
    &\quad \left. 1/(\tau + s_k - \ln(1-\rho_k^{(2)})) \right).
    \label{rk1}
\end{aligned}
\end{equation}

\noindent \textbf{Sampling transition matrix.} 
With $\bm{r}$ is marginalised out, we assume that $(l_{1k}^{(t)}, \cdots,l_{kk}^{(t)}) \sim \text{Mult} (l_{.k}^{(t)}, (\pi_{1k},\cdots, \pi_{Kk}))$. Via Dirichlet-Multinomial conjugacy, we have

\begin{align}
    (\pi_k|-) &\sim \text{Dir} (\alpha_{1k} + \sum_{t=1}^{T}l_{1k}^{(t)},  \cdots, \alpha_{Kk} + \sum_{t=1}^{T}l_{Kk}^{(t)}).
    \label{pi}
\end{align}

\noindent \textbf{Sampling binary community link $z_{k_1k_2}$ and its weights $w_{k_1k_2}$.}
We introduce an auxiliary variable $q_k$ where
\begin{equation}
    q_k \sim \text{Beta} (l_{.k}^{(.)}, \alpha_{.k}).
    \label{qk}
\end{equation}
Then we obtain 

\begin{equation}
    l^{(.)}_{kk} \sim \text{NB} (\alpha_{k_1k_2}, q_k).
\end{equation}
We further introduce $h_{k_1k_2} \sim \text{CRT} (l^{(.)}_{kk}, \alpha_{k_1k_2})$, where $\alpha = \text{Pr} (z_{k_1k_2})$, so that we can rewrite $h_{k_1k_2}$  as 

\begin{equation}
    h_{k_1k_2} \sim \text{Pois} (-z_{k_1k_2}\ln{(1-q_{k_2})}).
\end{equation}
Hence, we can obtain the condition distribution of $z_{k_1k_2}$ according to the Bayes' theorem as 

\begin{equation}
\begin{aligned}
    &(z_{k_1k_2}|-) \sim \text{Ber} (\frac{p_{k_1k_2}=1}{p_{k_1k_2}=1 + {p_{k_1k_2}=0}}), \text{where} \\
    &p_{k_1k_2}=1 \propto \text{Pr}(z_{k_1k_2} = 1){Pr}(h_{k_1k_2}|z_{k_1k_2} = 1), \\
    &p_{k_1k_2}=0 \propto \text{Pr}(z_{k_1k_2} = 0){Pr}(h_{k_1k_2}|z_{k_1k_2} = 0). \\
    \label{zkk}
\end{aligned}
\end{equation}

We sample the weights of binary community link as 
\begin{equation}
    (w_{k_1k_2}|-) \sim \text{Gam} (e_0 + \sum h_{k_1k_2}, e_0 - z_{k_1k_2}\ln{(1-q_k)}).
    \label{w_for_c}
\end{equation}

\noindent \textbf{Sampling latent subcounts in community links $\omega_{k_1k_2}$.}
We sample the latent subcount for each latent community links as 
\begin{equation}
    (\omega_{k_1k_2}|-) \sim  z_{k_1k_2}\text{Po}_{+}(\sum_{d_1=1}^{D} \sum_{d_2=1}^{D} m_{k_1d_1} v_{d_1d_2} m_{k_2d_2}),
    \label{wkk}
\end{equation}

\begin{equation}
    (\omega_{k_1d_1d_2k_2}|-) \sim \text{Mult} (\omega_{k_1k_2},(\frac{m_{k_1d_1} v_{d_1d_2} m_{k_2d_2}}{\sum_{d_1, d_2}^{D} m_{k_1d_1} v_{d_1d_2} m_{k_2d_2}})).  
    \label{wkddk}
\end{equation}

\noindent \textbf{Sampling membership of community and hierarchical community $m_{kd}$.} 
We can decompose $\omega_{k_1d_1k_2d_2}$ as $(\sum_{k_2d_2} \omega_{kd..})$, so we can sample $\omega_{k_1d_1k_2d_2}$ as 
\begin{equation}
    (\omega_{kd..}|-) \sim \text{Pois}(m_{k_1d_1} \sum_{k_2d_2}v_{d_1d_2} m_{k_2d_2}).
\end{equation}
Let $p_{kd} = \sum_{k_2d_2}v_{d_1d_2} m_{k_2d_2}$, via gamma-Poisson conjugacy, we obtain

\begin{equation}
    (m_{kd}|-) \sim \text{Gam}(a_k + \omega_{kd..}, 1/(c_k + p_{kd})).
    \label{mkd}
\end{equation}

\noindent \textbf{Sampling parameters related to $m_kd$.}
Firstly, accorrding to \emph{Lemma 2}, we can rewrite $\omega_{kd..} \sim \text{NB} (a_k, \frac{p_{kd}}{p_{kd} + c_k})$.
Then we introduce $l_{kd} \sim \text{CRT} (\omega_{kd..}, a_k)$ and base on \emph{Lemma 3}, we can aslo rewrite $l_{kd} \sim \text{Pois} (a_k \ln{(1+\frac{p_{kd}}{c_k})})$. Via Poisson-gamma conjugacy, we get

\begin{equation}
    (a_k|-) \sim \text{Gam} (e_0 + \sum_d l_{kd}, 1/(j_0 + \sum_d \ln{(1+\frac{p_{kd}}{c_k})}))
    \label{ak}
\end{equation}

We can directly use gamma-gamma conjugacy to sample $c_k$ as

\begin{equation}
    (c_k|-) \sim \text{Gam} (1+a_k, 1/(1+\sum_d m_{kd}))
    \label{ck}
\end{equation}

\noindent \textbf{Sampling hierarchical community matrix $v_{d_1d_2}$ and its weights $\lambda_d$.}
Since $\omega_{.d_1d_2.} = \sum_{k_1k_2} \omega_{k_1d_1d_2k_2}$, we can get $\omega_{.d_1d_2.} \sim \text{Pois} (v_{d_1d_2} \theta_{d_1d_2})$,  where $\theta_{d_1d_2} = 2^{-\delta_{k_1k_2}} \sum_{k_1k_2} m_{k_1d_1} m{k_2d_2}$. Here we denote 
that $\delta_{k_1k_2} = 1$ when $k_1 = k_2$. So we can obtain

\begin{equation}
    (v_{d_1d_2}|-) \sim \text{Gam} (\lambda_{d_1} \xi^{\delta_{k_1k_2}} \lambda_{d_2}^{1-\delta_{k_1k_2}} + \omega_{.d_1d_2.}, 1/(\beta + \theta_{d_1d_2})).
    \label{nudd}
\end{equation}

Using the same data augmentation method, we introduce $l_{d_1d_2} \sim \text{CRT} (\omega_{.d_1d_2.}, \lambda_{d_1}\xi^{\delta}\lambda_{d_2}^{1-\delta})$ and rewrite it as $l_{d_1d_2} \sim \text{Pois} (\lambda_{d_1}\xi^{\delta_{k_1k_2}} \lambda_{d_2}^{1-\delta_{k_1k_2}} \ln{(1+\frac{\theta_{d_1d_2}}{\beta})})$, so that we obtain

\begin{equation}
\begin{aligned}
    &(\lambda_{d_1}|-) \sim \text{Gam} (\gamma_1/D + \sum_{d_2}l_{d_1d_2}, \\
    &1/(c_0 + \sum_{d_2} (\xi^{\delta_{k_1k_2}} \lambda_{d_2}^{1-\delta_{k_1k_2}} \ln{(1+\frac{\theta_{d_1d_2}}{\beta})} ))),
    \label{lambda}
\end{aligned} 
\end{equation}

\begin{equation}
    (\xi|-) \sim \text{Gam} (e_0 + \sum_d l_{dd}, 1/(j_0 + \sum_d(\lambda_d \ln{(1+\frac{\theta_{dd}}{\beta})}))).
    \label{xi}
\end{equation}

\noindent \textbf{Sampling parameters related to $\lambda_d$.}
We introduce $\hat{l_d} \sim \text{CRT} (l_{d_{1}}, \gamma_1 /D)$, where $l_{d_{1}} = \sum_{d_{2}} l_{d_1d_2}$, and we get 

\begin{equation}
    (\gamma_1|-) \sim \text{Gam} (1+\sum \hat{l_d}, 1/(1+ \frac{1}{D}\ln{(1+\hat{p_k})})),
    \label{gamma1}
\end{equation}
where $\hat{p_k} = \frac{\sum_{d_2} (\xi^{\delta_{k_1k_2}} \lambda_{d_2}^{1-\delta_{k_1k_2}} \ln{(1+\frac{\theta_{d_1d_2}}{\beta})} )}{c_0 + \sum_{d_2} (\xi^{\delta_{k_1k_2}} \lambda_{d_2}^{1-\delta_{k_1k_2}} \ln{(1+\frac{\theta_{d_1d_2}}{\beta})} )}$.
Via gamma-gamma conjugacy,

\begin{equation}
    (c_0|-) \sim \text{Gam} (1 + \gamma_1 /D, 1/(1+\sum_d \lambda_d)).
    \label{c0}
\end{equation}

Algorithm \ref{algorithm} summarises the full sampling procedure of G-HSEPM.
\begin{algorithm}[!ht]
    \renewcommand{\algorithmicrequire}{\textbf{Input:}}
	\renewcommand{\algorithmicensure}{\textbf{Output:}}
	\caption{Gibbs sampling algorithm for G-HSEPM}
    \begin{algorithmic}[1] 
        \REQUIRE  dynamic relational data $B^{(1)}, \cdots, B^{(T)}$.  \\
                    Initialise the number of communities $K$, the number of hierarchical communities $D$ and other parameters; 
        \REPEAT
            \STATE Sample $x_{ij}^{(t)}$ (Eq.\ref{xij}) for non-zeros links and update $x_{ijk}^{(t)}$ (Eq.\ref{xijk})
            \STATE Sample $\phi_{ik}$ (Eq.\ref{phi}) 
            
            \FOR {$t=T,\cdots,1$}
                \IF{$t = T$}
                    \STATE Sample $l_{k}^{(T)}$ (Eq.\ref{lTk})
                \ELSIF{$t = T-1, \cdots, 2$}
                    \STATE Sample $l_{k}^{(t)}$ (Eq.\ref{ltk})
                \ELSIF{$t = 1$}
                    \STATE Sample $l_{k}^{(1)}$ (Eq.\ref{l1k})
                \ENDIF               
                \STATE Update $\rho_k^{(t)}$ (Eq.\ref{rho})
            \ENDFOR
            
            \FOR {$t=1,\cdots,T$}
                \IF{$t = 1$}
                    \STATE Sample $r_k^{(1)}$ (Eq.\ref{rk1})
                \ELSIF{$t = 2,\cdots,T-1$}
                    \STATE Sample $r_{k}^{(t)}$ (Eq.\ref{rkt})
                \ELSIF{$t = T$}
                    \STATE Sample $r_{k}^{(T)}$ (Eq.\ref{rkT})
                \ENDIF
            \ENDFOR

            \STATE Sample $\pi_k$ (Eq.\ref{pi}), update $q_k$ (Eq.\ref{qk}), $w_{k_1k_2}$ (Eq.\ref{w_for_c}) and calculate $z_{k_1k_2}$ (Eq.\ref{zkk})

            \STATE Sample $\omega_{k_1k_2}$ (Eq.\ref{wkk}) and update $\omega_{k_1d_1d_2k_2}$ (Eq.\ref{wkddk})
            \STATE Sample $m_{kd}$(Eq.\ref{mkd})
            \STATE Sample $v_{d_1d_2}$(Eq.\ref{nudd})
            \STATE Sample $\gamma_1$(Eq.\ref{gamma1}) and $c_0$(Eq.\ref{c0})
            
        \UNTIL convergence;

        \ENSURE Posterior mean of $\phi_{ik}$ and $r_k^{(t)}$ 
    \end{algorithmic}
    \label{algorithm}
\end{algorithm}

\section{Gibbs sampling for HSEPM}
\label{Gibbs_HSEPM}
We shall first demonstrate the structure of HSEPM, and then present the sampling procedure of this model. 

\begin{figure}[hbt!]
    \centering
    \includegraphics[width=0.5\columnwidth]{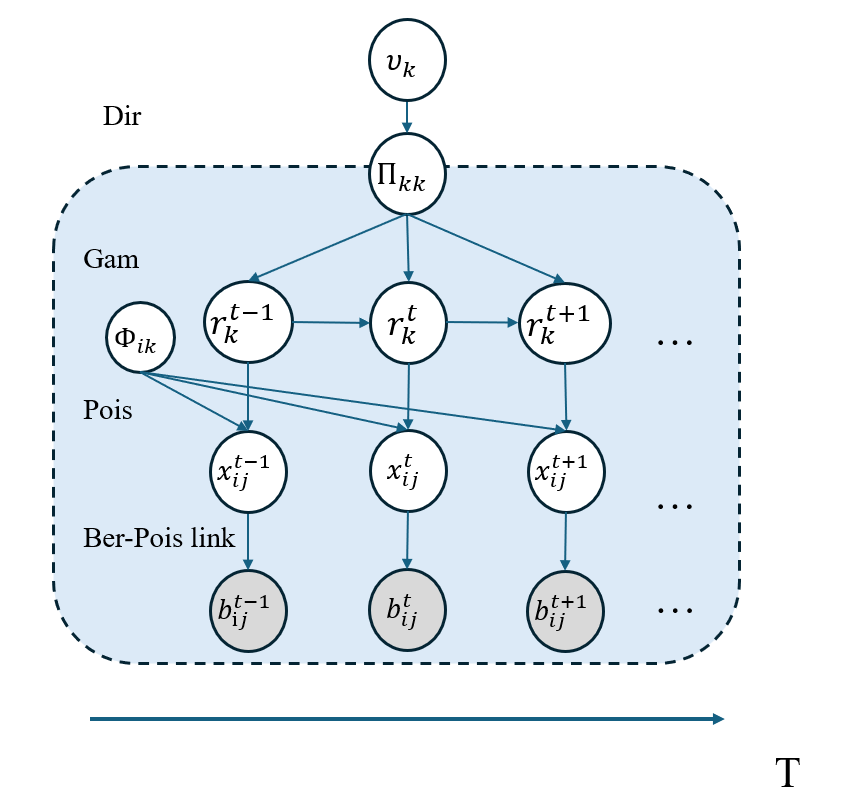}
    \caption{The graphical structure of the HSEPM model.}
    \label{Modelstructure_HSEPM}
\end{figure}

Figure \ref{Modelstructure_HSEPM} shows the graphical diagram of the HSEPM model, where the transition kernel is controlled by a parameter $\bm{\nu}$, and the remaining structure is similar to the G-HSEPM model.  
For parameter inference, we also use Gibbs sampling as follows. 

\noindent \textbf{Sampling latent counts $x_{ij}^{(t)}$.} As similar to G-HSEPM, we sample the latent count for each observed edge $b_{ij}^{(t)}$ at each time step as

\begin{equation}
    (x_{ij}^{(t)}|-) \sim b_{ij}^{(t)}\text{Po}_{+}(\sum_{k=1}^{K}r_k^{(t)}\phi_{ik}\phi_{jk}).
    \label{xij2}
\end{equation}
and the latent count $x_{ijk}^{(t)}$ can be sampled as

\begin{equation}
    (x_{ijk}^{(t)}|-) \sim \text{Mult} (x_{ij}^{(t)},(\frac{r_k^{(t)}\phi_{ik}\phi_{jk}}{\sum_{k=1}^K r_k^{(t)}\phi_{ik}\phi_{jk}})).
    \label{xijk2}
\end{equation}

\noindent \textbf{Sampling membership of nodes and communities $\phi_{ik}$.} Via the gamma-Poisson conjugacy, we obtain 

\begin{equation}
    (\phi_{ik}|-) \sim \text{Gam} (a_0 + \sum_{t=1}^{T} \sum_{j\neq i}^{N} x_{ijk}^{(t)}, 1/(c_i + \sum_{t=1}^{T} \sum_{j\neq i}^{N} r_{k}^{(t)}\phi_{jk})).
    \label{phi2}
\end{equation}



\noindent \textbf{Sampling community weights $r_{k}^{(t)}$.} 
For $t = T$,  $x_{..k}^{(T)} = \sum_{i}^{N-1} \sum_{j=(i+1)}^{N} x_{ijk}^{(T)}$, $x_{..k}^{(T)} \sim \text{Pois} (r_k^{(t)}s_k)$, where $s_k = \sum_{i}^{N-1} \sum_{j=(i+1)}^{N} \phi_{ik}\phi_{jk} $. 

To obtain the backward information, we use augmentation techniques and introduce an auxiliary variable $l_k$, which is sampled as  
\begin{equation}
    l_{.k}^{(T)} \sim \text{CRT} (x_{..k}^{(T)}, \sum_{k_2=1}^{K} \pi_{kk_2} r_{k_2}^{(T-1)}),
    \label{lkT2}
\end{equation}
when $t = T$,
\begin{equation}
    l_{.k}^{(t)}\sim\text{CRT} (x_{..k}^{(t)} + l_{.k}^{(t+1)}, \sum_{k_2=1}^{K} \pi_{kk_2} r_{k_2}^{(t-1)}),
    \label{lkt2}
\end{equation}
when $t = T-1, \cdots, 2$, and 
\begin{equation}
    l_{k}^{(1)} \sim \text{CRT} (x_{..k}^{(1)}+l_{.k}^{(2)}, \tau\nu), 
    \label{lk12}
\end{equation}
when $t = 1$.

To obtain the forward information, we sample $r_k^{(t)}$ as 
\begin{equation}
\begin{aligned}
    (r_k^{(1)}|-) &\sim \text{Gam} \left( x_{..k}^{(1)} + l_{.k}^{(2)} + \tau\nu, \right. \\
    &\quad \left. 1/(\tau + s_k - \ln(1-\rho_k^{(2)})) \right),
    \label{rk12}
\end{aligned}
\end{equation}
when $t = 1$, 
\begin{equation}
\begin{aligned}
    (r_k^{(t)}|-) &\sim \text{Gam} \left( x_{..k}^{(t)} + l_{.k}^{(t+1)} + \sum_{k_{2}}^{K}\pi_{kk_{2}}r_{k_{2}}^{(t-1))}, \right. \\
    &\quad \left. 1/(\tau + s_k - \ln(1-\rho_k^{(t+1)})) \right),
    \label{rkt2}
\end{aligned}
\end{equation}
when $t = 2, \cdots, T-1$, and 
\begin{equation}
    (r_k^{(T)}|-) \sim \text{Gam} (x_{..k}^{(T)} + \sum_{k_2=1}^{K} \pi_{kk_2} r_{k_2}^{(T-1)}, 1/(\tau + s_k)),
    \label{rkT2}
\end{equation}
when $t = T$, where 
\begin{equation}
    \rho_k^{(t+1)}=\frac{s_k-\ln{(1-\rho_k^{(t+2)})}}{\tau+s_k-\ln{(1-\rho_k^{(t+2)})}}.
    \label{rho2}
\end{equation}

\noindent \textbf{Sampling transition matrix $\Pi$.}
With $r$ is marginalised out, assumes that $(l_{1k}^{(t)}, \cdots,l_{kk}^{(t)}) \sim \text{Mult} (l_{.k}^{(t)}, (\pi_{1k},\cdots, \pi_{Kk}))$. Via Dirichlet-Multinomial conjugacy, 

\begin{align}
    (\pi_k|-) &\sim \text{Dir} (\nu_1\nu_k + \sum_{t=1}^{T}l_{1k}^{(t)}, \cdots, \xi\nu_k + \sum_{t=1}^{T}l_{kk}^{(t)}, \cdots, \notag \\
    &\quad \nu_K\nu_k + \sum_{t=1}^{T}l_{Kk}^{(t)}).
    \label{pi2}
\end{align}

\noindent \textbf{Sampling $\nu$ and $\xi$.}
Firstly, we marginalise over $\Pi$ to obtain a Dirichlet-Multinomial distribution

\begin{equation}
    (l_{1k}^{(.)},\cdots,l_{Kk}^{(.)}) \sim \text{DirMult} (l_{.k}^{(.)}, (\nu_1\nu_k,\cdots,\xi\nu_k,\cdots,\nu_K\nu_k)).
\end{equation}
The Dirichlet-Multinomial distribution can convert to the negative binomial distribution when introducing the following beta-distributed auxiliary variable \cite{zhou2018nonparametric},

\begin{equation}
    q_k \sim \text{Beta} (l_{.k}^{(.)}, \nu_k(\xi + \sum_{k_{1}\neq k} \nu_{k_{1}})),
    \label{qk2}
\end{equation}
and we get $l_{kk}^{(.)} \sim \text{NB} (\xi\nu_k, q_k)$ and $l_{k_{1}k}^{(.)} \sim \text{NB} (\nu_{k_{1}}\nu_{k},q_k)$.  We further introduce a auxiliary variables $h_{k_1k}$ for $k_1 \neq k$ as 

\begin{equation}
    \begin{aligned}
        h_{kk} &\sim \text{CRT}(l_{kk}^{(.)}, \xi\nu_k), \\
        h_{k_1k} &\sim \text{CRT}(l_{k_{1}k}^{(.)}, \nu_{k_{1}}\nu_k).
    \end{aligned}
    \label{hkk2}
\end{equation}
Similar to the augmentation and sampling method in sampling $r_k^{(t)}$, $\xi$ can be sampled as
\begin{equation}
    (\xi|-) \sim \text{Gam} (f_0+\sum_{(k=1)}^{K}h_{kk}, 1/(g_0 - \sum_{k=1}^{K}\nu_k\ln{(1-q_k)})).
    \label{xi2}
\end{equation}
Next, we introduce 
\begin{equation}
    n_k = h_{kk} + \sum_{k_1 \neq k}h_{k_{1}k} + \sum_{k_2 \neq k}h_{kk_{2}} + l_{k.}^{(1)},
\end{equation}
where $l_{k.}^{(1)} \sim \text{Pois} (\tau\nu_k\ln{(1-\rho_k^{(1)})})$. Using the Poisson additive property and gamma-Poisson conjugacy, we have

\begin{equation}
    (\nu_k|-) \sim \text{Gam} (\frac{\gamma_0}{K} + n_k, 1/(\beta + t_k)),
    \label{nu2}
\end{equation}
where 
\begin{equation}
\begin{split}
    t_k = & -\ln{(1-q_k)}(\xi+\sum_{k_{1} \neq k}\nu_{k_{1}}) \\
          & - \sum_{k_{2}\neq k} \ln{(1-q_{k_{2}}}) \nu_{k_{2}}  + \ln{(1-\rho_k^{(1)})} \tau.
\end{split}
\end{equation}

\noindent \textbf{Sampling $\beta$.}
Via gamma-gamma conjugacy, we obtain
\begin{equation}
    (\beta|-) \sim \text{Gam} (f_0+\gamma_0, 1/(g_0 + \sum_{k=1}^{K}\nu_k)).
    \label{beta2}
\end{equation}

Algorithm \ref{algorithm2} summarises the full sampling procedure of HSEPM.
\begin{algorithm}[!ht]
    \renewcommand{\algorithmicrequire}{\textbf{Input:}}
	\renewcommand{\algorithmicensure}{\textbf{Output:}}
	\caption{Gibbs sampling algorithm for HSEPM}
    \begin{algorithmic}[1] 
        \REQUIRE  dynamic relational data $B^{(1)}, \cdots, B^{(T)}$.  \\
                    Initialise the number of communities $K$ and other parameters; 
        \REPEAT
            \STATE Sample $x_{ij}^{(t)}$ (Eq.\ref{xij2}) and update $x_{ijk}^{(t)}$ (Eq.\ref{xijk2})
            \STATE Sample $\phi_{ik}$ (Eq.\ref{phi2})
            
            \FOR {$t=T,\cdots,1$}
                \IF{$t = T$}
                    \STATE Sample $l_{k}^{(T)}$ (Eq.\ref{lkT2})
                \ELSIF{$t = T-1, \cdots, 2$}
                    \STATE Sample $l_{k}^{(t)}$ (Eq.\ref{lkt2})
                \ELSIF{$t = 1$}
                    \STATE Sample $l_{k}^{(1)}$ (Eq.\ref{lk12})
                \ENDIF               
                \STATE Update $\rho_k^{(t)}$ (Eq.\ref{rho2})
            \ENDFOR
            
            \FOR {$t=1,\cdots,T$}
                \IF{$t = 1$}
                    \STATE Sample $r_k^{(1)}$ (Eq.\ref{rk12})
                \ELSIF{$t = 2, \cdots, T-1$}
                    \STATE Sample $r_k^{(t)}$ (Eq.\ref{rkt2})
                \ELSIF{$t = T$}
                    \STATE Sample $r_{k}^{(T)}$ (Eq.\ref{rkT2})
                \ENDIF
            \ENDFOR

            \STATE Sample $\pi_k$ (Eq.\ref{pi2}) update $q_k$ (Eq.\ref{qk2}) and $h_{kk}$ (Eq.\ref{hkk2})

            \STATE Sample $\xi$(Eq.\ref{xi2} and $\nu_{k}$(Eq.\ref{nu2})
            \STATE Sample $\beta$(Eq.\ref{beta2})
            
        \UNTIL convergence;

        \ENSURE Posterior mean of $\phi_{ik}$ and $r_k^{(t)}$ 
    \end{algorithmic}
    \label{algorithm2}
\end{algorithm}

\section{Additional Results}
\label{Additional Results}

\begin{figure}[ht]
    \centering
    \begin{subfigure}[b]{0.7\columnwidth}
        \includegraphics[width=\columnwidth]{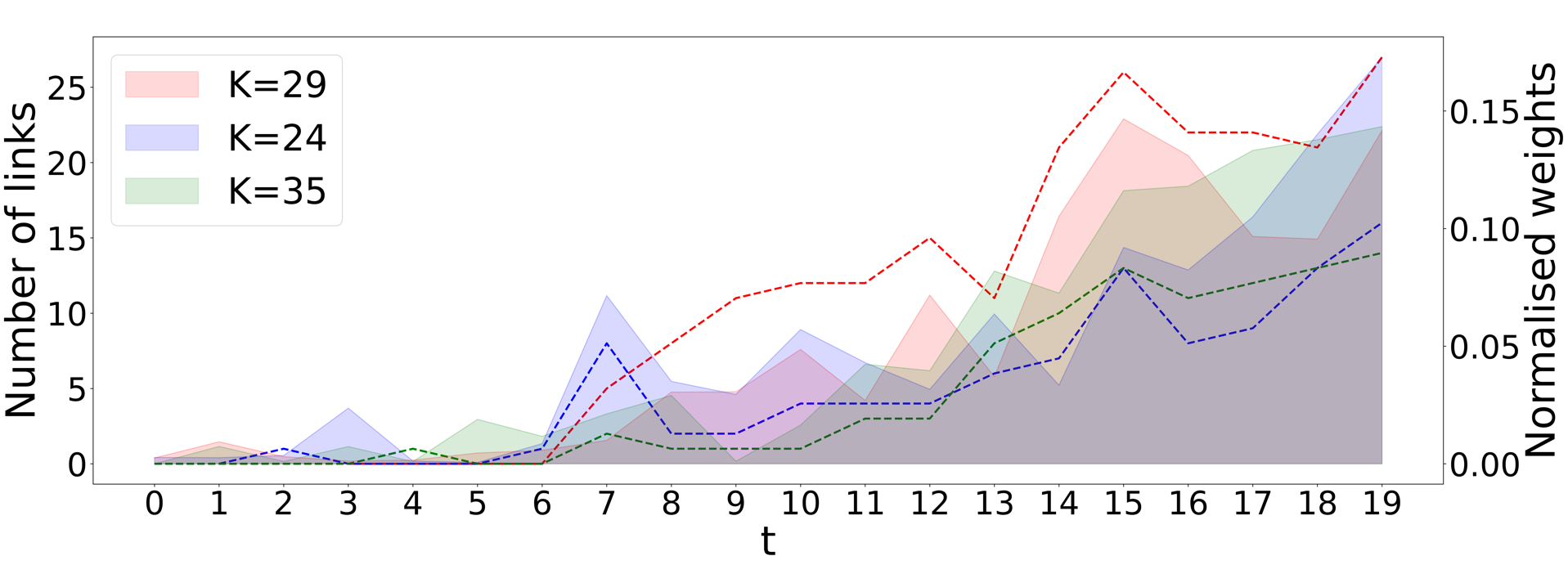}
        \caption{}
    \end{subfigure} 
    \vfill
    \begin{subfigure}[b]{0.7\columnwidth}
        \includegraphics[width=\columnwidth]{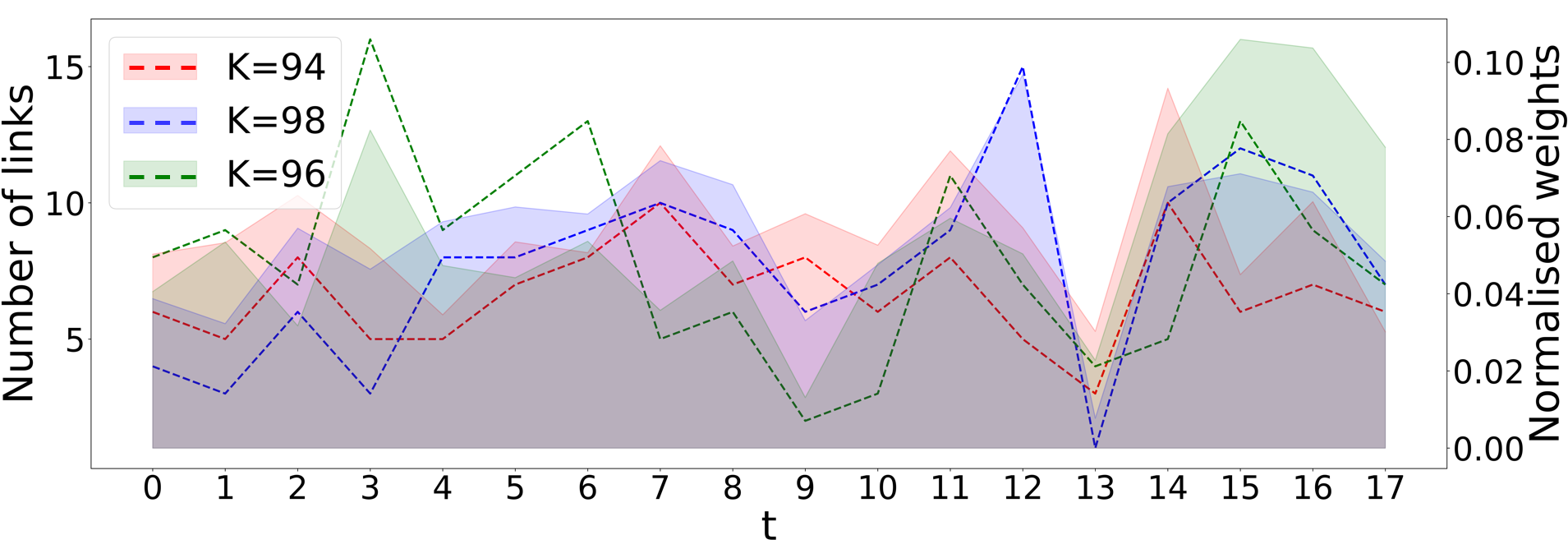}
        \caption{}
    \end{subfigure}
    \vfill
    \begin{subfigure}[b]{0.7\columnwidth}
        \includegraphics[width=\columnwidth]{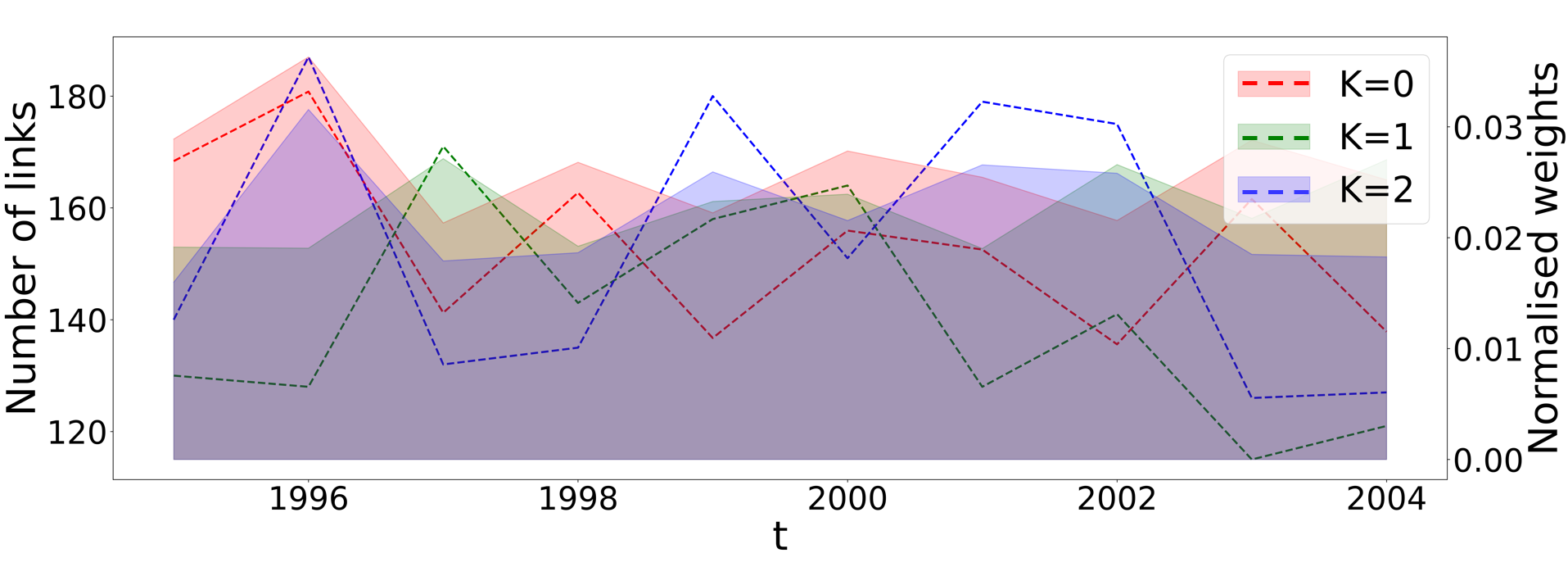}
        \caption{}
    \end{subfigure}

    \caption{Plot (a), (b) and (c) show the evolution of link counts among major vertices and the normalised weights of the top 3 communities over time in the Enron, EuEmail and DBLP datasets, respectively. The red, green and blue colours represent different communities. Dotted lines indicate the number of links, while the colour-filled  areas depict the normalised community weights.}      
    \label{NumLinks_and_Weights}
\end{figure}

Figure \ref{NumLinks_and_Weights} illustrates the correlation between community evolution and vertex interactions in the Enron, emailEu, and DBLP datasets. The corresponding plot for the vdBunt dataset is provided in the main text (Section \ref{Dynamic community detection}).

\begin{figure}[hbt!]
    \centering
    \begin{subfigure}[b]{0.2\columnwidth}
        \includegraphics[width=\columnwidth]{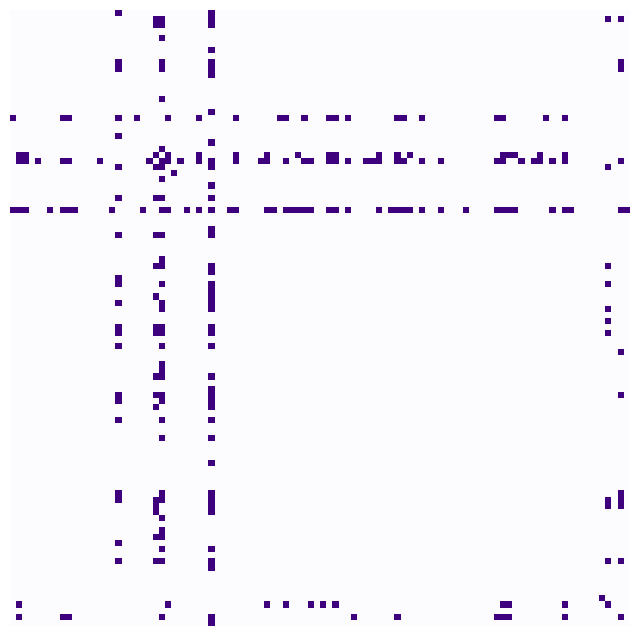}
    \end{subfigure} 
    \hspace{1.3cm}
    \begin{subfigure}[b]{0.2\columnwidth}
        \includegraphics[width=\columnwidth]{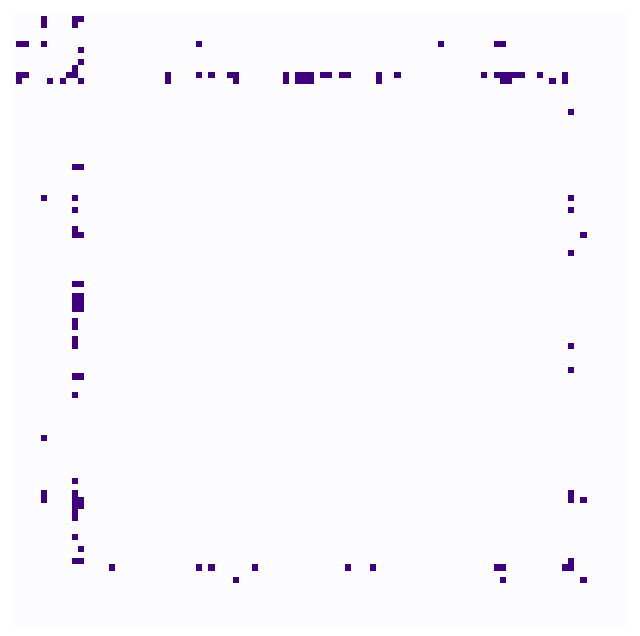}
    \end{subfigure}
    \vfill
    \begin{subfigure}[b]{0.32\columnwidth}
        \includegraphics[width=\columnwidth]{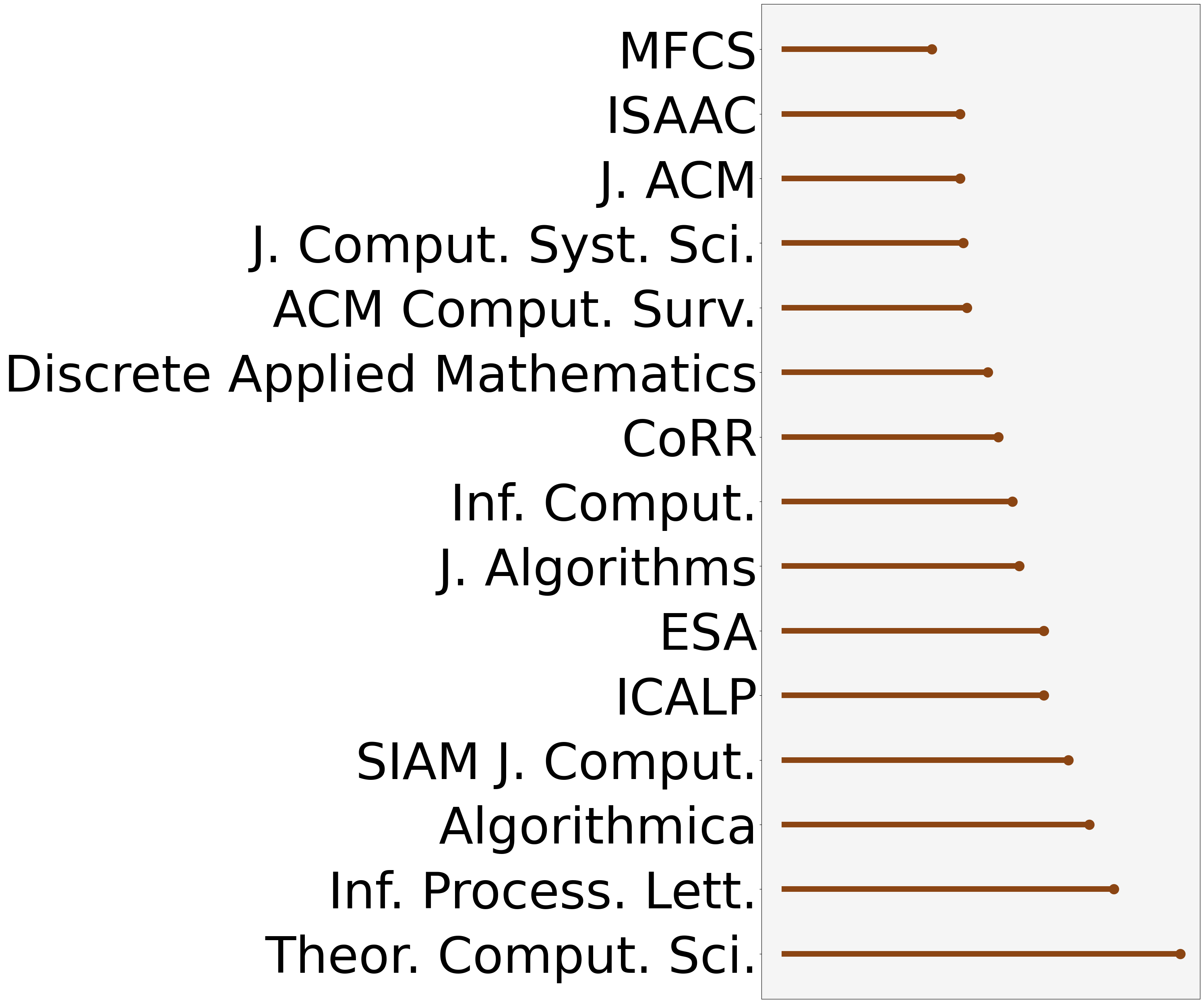}
        \caption{}
    \end{subfigure}
    \hspace{0.1cm}
    \begin{subfigure}[b]{0.32\columnwidth}
        \includegraphics[width=\columnwidth]{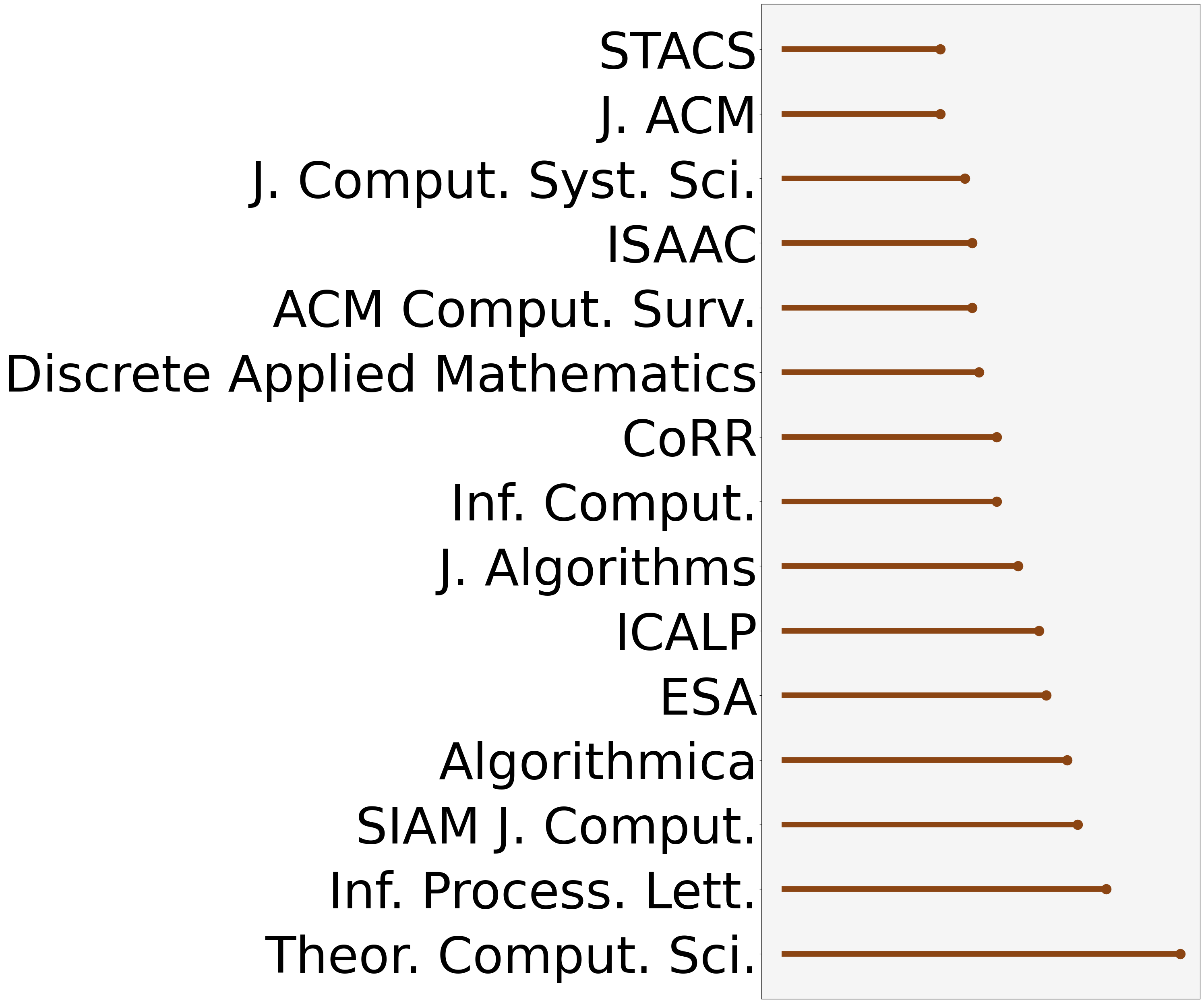}
        \caption{}
    \end{subfigure}
    
    \hspace{1.3cm}
    \begin{subfigure}[b]{0.2\columnwidth}
        \includegraphics[width=\columnwidth]{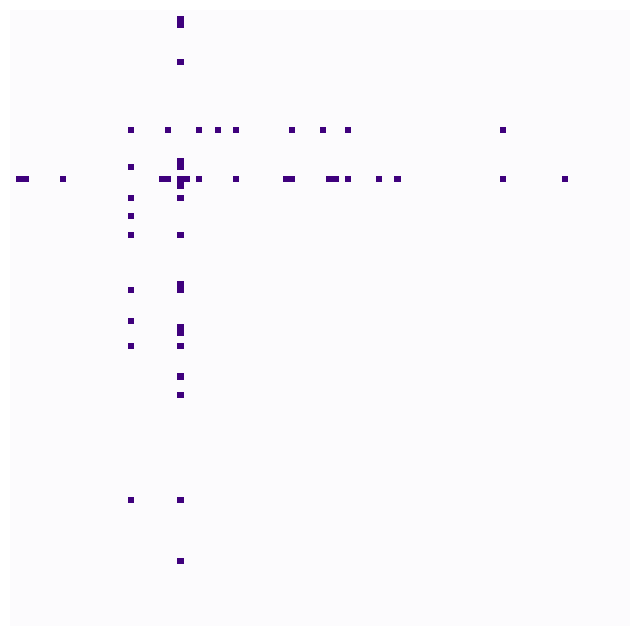}
    \end{subfigure}
    \hspace{1.3cm}
    \begin{subfigure}[b]{0.2\columnwidth}
        \includegraphics[width=\columnwidth]{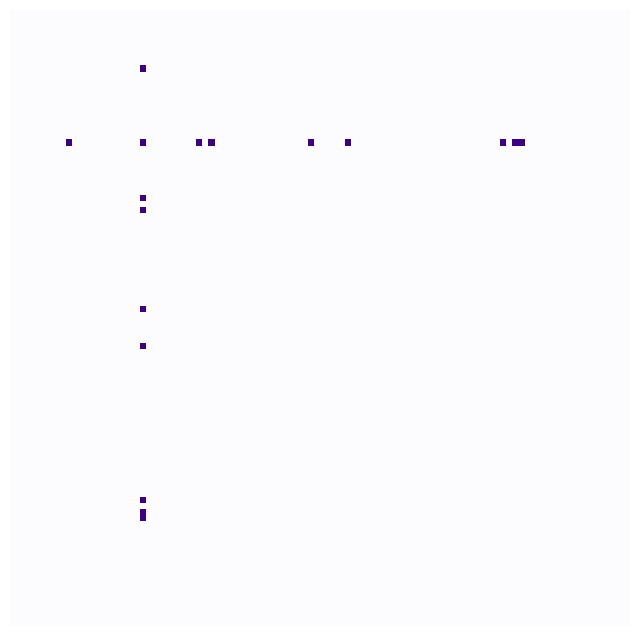}
    \end{subfigure}
    \vfill
    \begin{subfigure}[b]{0.32\columnwidth}
        \includegraphics[width=\columnwidth]{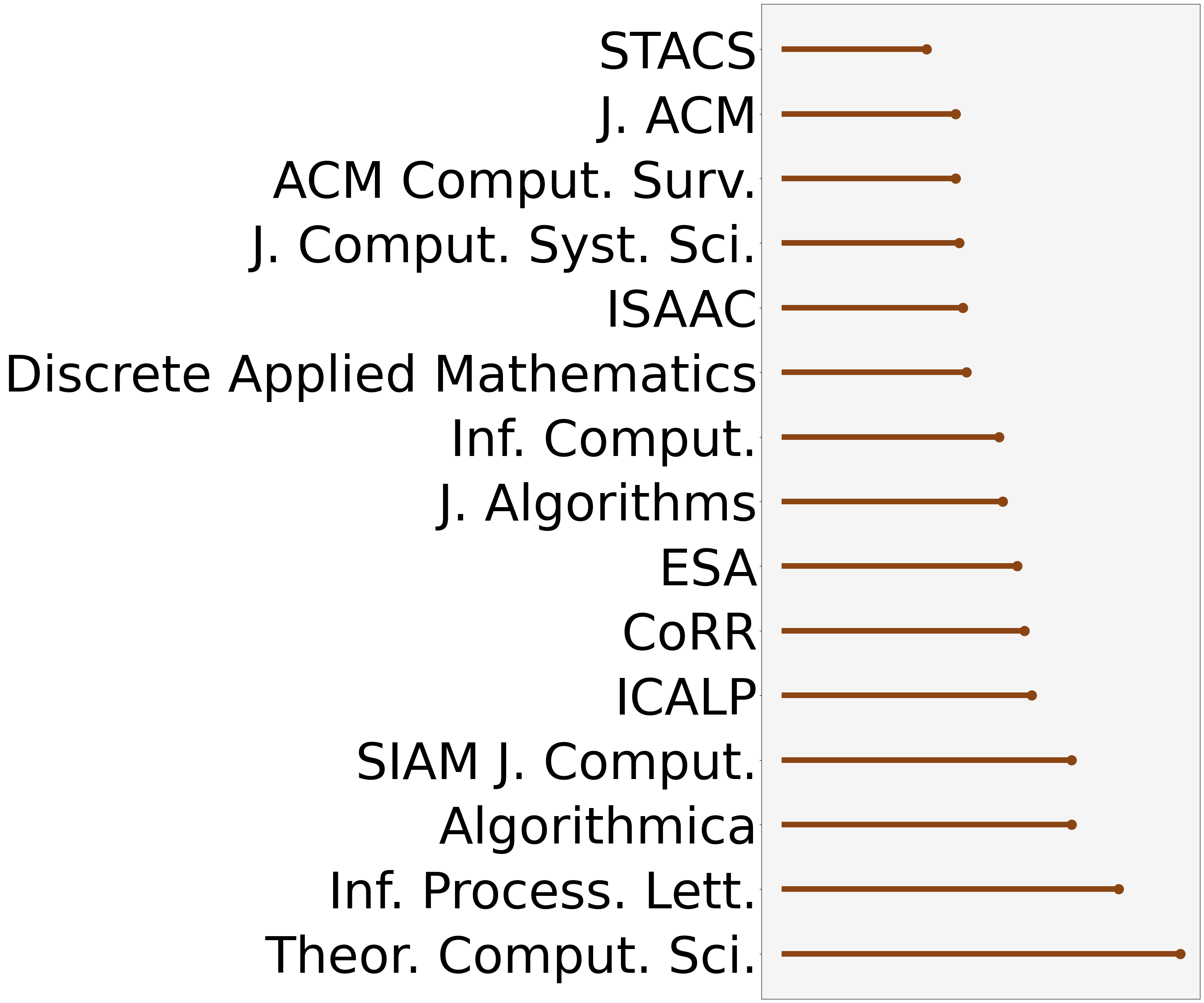}
        \caption{}
    \end{subfigure}
    \hspace{0.1cm}
    \begin{subfigure}[b]{0.32\columnwidth}
        \includegraphics[width=\columnwidth]{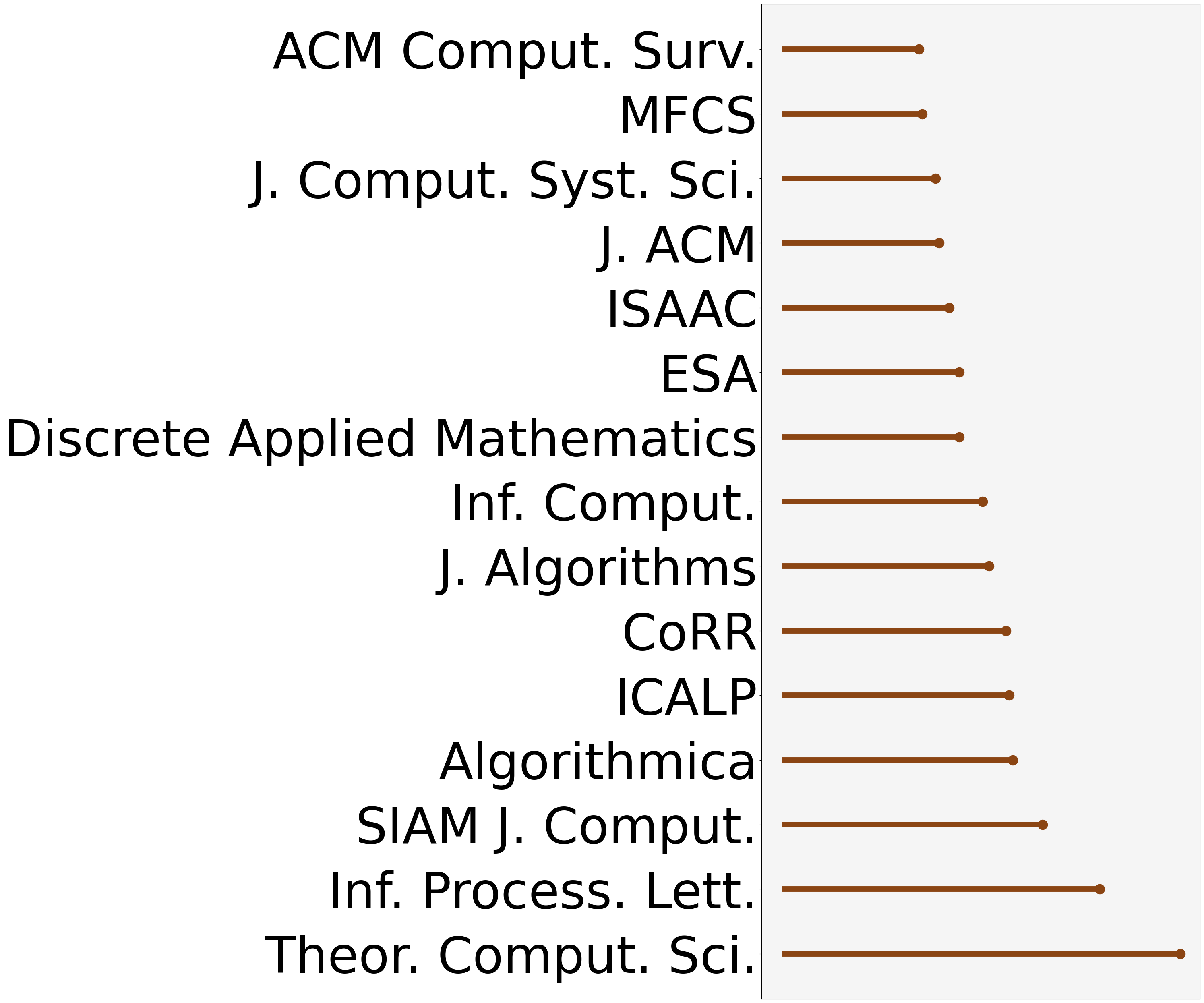}
        \caption{}
    \end{subfigure}

    \caption{Plot (a), (b), (c) and (d) represent the hierarchical communities beyond the top 3 hierarchical communities as discussed in the main text. The plots on the top show the relationship between communities in each hierarchical community. The plots on the bottom illustrate the publication venues where these communities published most frequently over the past 10 years. Although, as discussed in the main text, they share some common research areas such as theoretical computer science and algorithm design, there are also some differences: (a) focuses more on conferences related to computing and algorithms, (b) emphasises foundational theories of computer science, (c) is less oriented toward mathematics-driven research, while (d) focuses more on research with a strong mathematical foundation.} 
    \label{DBLP_Journal_S}
\end{figure}

Figure \ref{DBLP_Journal_S} provides additional information of the latent hierarchical communities on the DBLP dataset as analysed in the main text (Section \ref{Exploratory analysis}).

\end{document}